\documentclass[journal=jacsat,manuscript=article]{achemso}

\usepackage[version=3]{mhchem} 
\usepackage[normalem]{ulem}
\usepackage{graphicx} 
\graphicspath{{Figs/}} 

\author{Ji Feng}
\author{Altai J. Perry}
\author{Xiaojing Weng}
\affiliation{Department of Mechanical Engineering, University of California at Riverside, Riverside, CA, 92521 USA}
\author{Alma K. González-Alcalde}
\affiliation{Department of Mechanical Engineering, University of California at Riverside, Riverside, CA, 92521 USA}
\alsoaffiliation{Current address: Departamento de Electrónica y Telecomunicaciones, Centro de Investigación Científica y de Educación Superior de Ensenada (CICESE), Ensenada, BC, México.}
\author{Oriol Arteaga}
\affiliation{Dep. Física Aplicada, Plat Group, IN2UB, Universitat de Barcelona, Barcelona 08028, Spain}
\author{Mario J. Mencagli}
\affiliation{Department of Electrical and Computer Engineering, University of Delaware, Newark, DE, 19716 USA}
\author{Luat T. Vuong}

\affiliation{Department of Mechanical Engineering, University of California at Riverside, Riverside, CA, 92521 USA}
\email{LuatV@UCR.edu}
\title
  {Polarimetric
compressed sensing with hollow, self-assembled diffractive films}

\abbreviations{IR,NMR,UV}

\begin{document}


\begin{abstract}
Sensing light's polarization and wavefront direction enables surface curvature assessment, material identification, shadow differentiation, and improved image quality in turbid environments. Traditional polarization cameras utilize multiple sensor measurements per pixel and polarization-filtering optics, which result in reduced image resolution. We propose a nanophotonic pipeline that enables compressive sensing and reduces the sampling requirements with a low-refractive-index, self-assembled optical encoder. These nanostructures scatter light into lattice modes, which encode the wavefront direction and the polarization ellipticity in the linearly-polarized components of the diffracted, interference patterns. Combining optical encoders with a neural network, the system predicts pointing and polarization when the interference patterns are adequately sampled. A comparison of ``ordered'' and ``random'' optical encoders shows that the latter both blurs the interference patterns and achieves higher resolution. Our work centers on the unexpected modulation and spatial multiplexing of incident light polarization by self-assembled hollow nanocavity arrays as a class of materials distinct from traditional metasurfaces that will not only enable encoding for polarization and optical computing but also for compressed sensing and imaging.
\end{abstract}

\section{Introduction}
The capacity to sense both light polarization and its wavefront or pointing direction offers the ability to infer the surface curvature, material, or texture, distinguish shadows from objects, see otherwise-transparent shapes, or improve image quality in turbid environments \cite{Lin_2006, Schechner_2003, Rowe_1995, Yu_2017}. However, as independent qualities of light, polarization and pointing are typically detected separately with arrays of sensors and polarization-filtering optics. Polarization states are characterized by the Jones or Stokes parameters, i.e., four separate readings that identify the power (${\rm S}_0$), 2-$D$ linear-polarization (LP) axis (${\rm S}_1$, ${\rm S}_2$), and circular-polarization (CP) ellipticity (${\rm S}_3$). Since {each polarization-filtered sensor provides a projection of the incoming wave, multiple sensor measurements are needed for each sampled pixel; ``full-Stokes'' cameras involve inevitable reductions in image resolution.

The process of physical encoding would mitigate the challenge of {{reduced resolution}} caused by polarization sensitivity through compressive sensing. Compressive sensing is a revolutionary technique that exploits the sparsity of a signal so that the data acquisition and storage requirements are significantly reduced without sacrificing the reconstruction quality \cite{Cand_s_2007}. By contrast, {conventional imaging is a} relatively cumbersome direct-sampling approach [Fig. \ref{fig:intro1}(a)]: {it} involves rasterized or serial measurements to identify pointing and polarization with polarization filters and multiple-pixel sensors. {With conventional imaging, the grid sampling dimension $\Lambda$ must be less than the diameter of the diffraction-limited spot \textit{d}, or $\Lambda < \textit{d}$. Moreover, the polarization inferred is generally only associated with the polarization that is measured.} A physical encoder that aids compressed sensing maps the input optical fields over different output field components in a manner analogous to {and generally paired with} computing algorithms {[Fig. \ref{fig:intro1}(b)]}. 

Consider the characterization of the wavefront or pointing direction, which typically involves locating the centroid of a hotspot from a serial readout of filtered detectors. The calculation can be written,
\begin{equation}
\mathbf{y=Ax},
\end{equation}
where the sensing matrix $\mathbf{A}$ transfers the vector of input measurements $\mathbf{x}$ to a $1\times 4$ sensor output $\mathbf{y}$: this sensing approach is inefficient because many measurements are needed and there are many zero-valued elements in the sensing matrix $\mathbf{A}$. These zero-valued elements signify that the algorithmic operations performed by $\mathbf{A}$ are inefficient and yield no information. 

\begin{figure*}[htb]
  \centering
  \includegraphics[width=.7\linewidth]{ 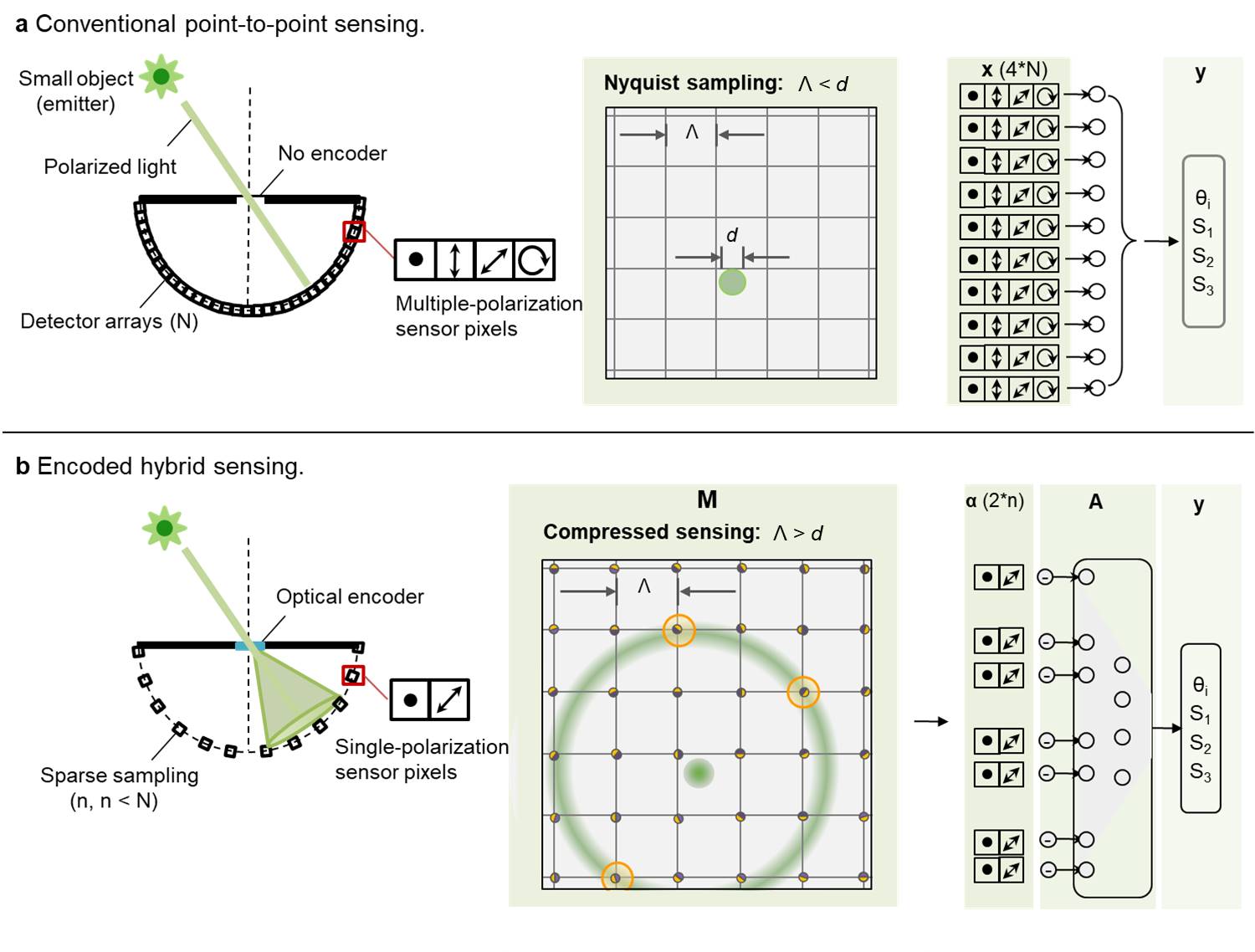}
  \caption{Basic schematic of conventional polarimetric imaging and proposed polarimetric compressed sensing. {For a diffraction-limited spot with diameter \textit{d}, the Nyquist sampling limit refers to the minimum sampling dimension needed to capture the spot size or a grid sample dimension of $\Lambda < \textit{d}$.} (a) With a conventional polarimetric camera, the direction and polarization of a point source are measured with independent and separate filters and pixels. Each image vector $\mathbf{x}$ contains many zero-valued readings. (b) With encoded and hybrid sensing, the point light source and its polarization are mapped to multiple pixels with different polarizations. Each encoded image vector $\mathbf{\alpha}$ carries joint statistics of the pointing and polarization.}
  \label{fig:intro1}
\end{figure*}
\begin{figure}[htb]
  \centering \includegraphics[width=\linewidth]{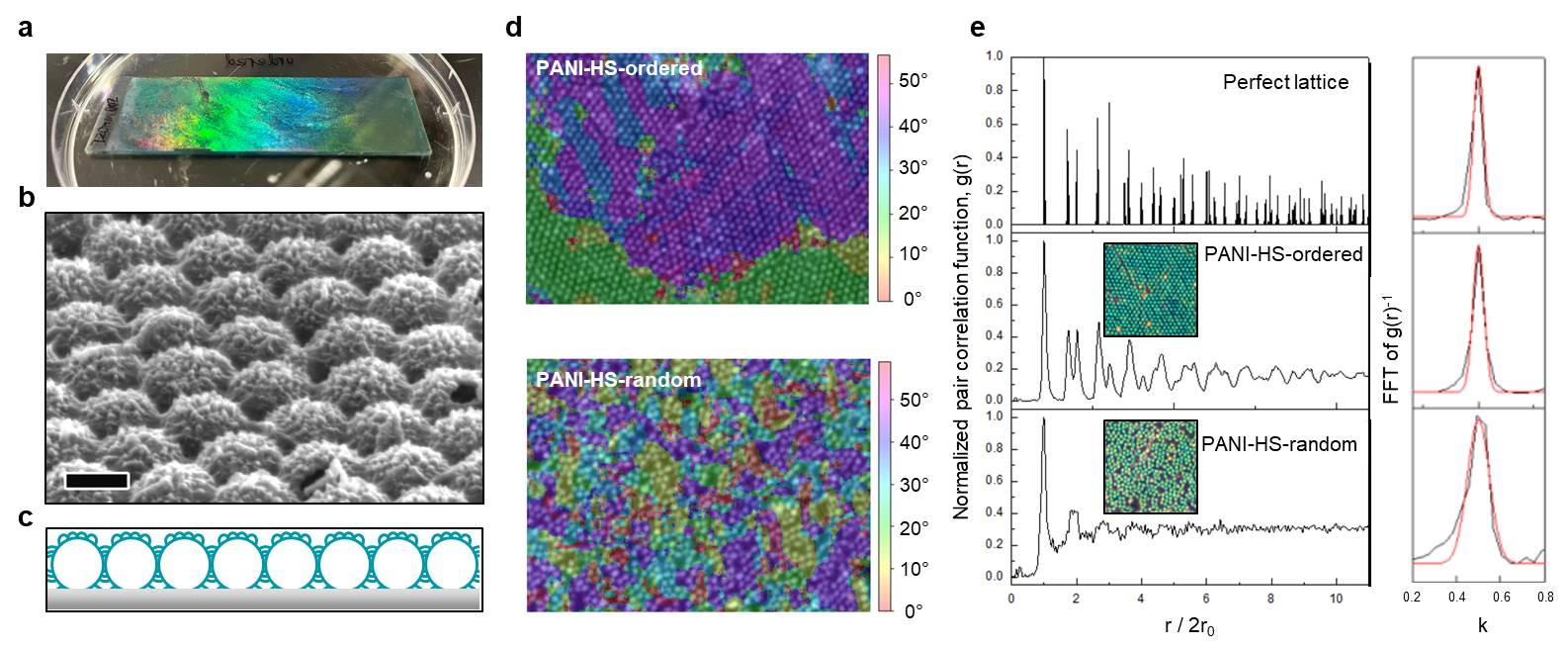}  \caption{Characterization of the optical encoders. Digital image (a) and SEM image (b) of the PANI-HS optical encoder. {Scale bar is 500 nm.} (c) Cartoon illustration showing the cross section of the PANI-HS optical encoder. (d) {Grain boundary} analyses for ``ordered'' (top) and ``random'' samples (bottom) showing the grain axes rotation (degrees). {Microscope images are 30 $\mu$m x 22 $\mu$m}. (e) The pair correlation function $g(r)$ computed from optical images against the normalized radius $r/2r_0$, where $r$ is the distance from an arbitrary origin, and $r_0$ is the mean sphere radius. The full width at half maximum (FWHM) of the first peak in the Fourier transform of the function $g(r)^{-1}$ provides a measure of the structural disorder. Perfect lattice (top), example ``ordered''(middle) and ``random'' sample (bottom) are shown. }\label{fig:intro2}
\end{figure}

The physical encoder offers a more efficient approach. It can be represented mathematically with an intermediate transformation $\mathbf{M}$, 
\begin{eqnarray}
\mathbf{ y} &=& \mathbf{\tilde{A}(Mx)}\\
&=& \mathbf {\tilde{A}\alpha} ,
\end{eqnarray}
that enables prediction of $\mathbf{y}$ with fewer measurements $\mathbf{\alpha}$. The effective sensing operation, $\mathbf{\tilde{A}}$, which now represents a procedure learned by an optimization algorithm, contains fewer zeros \cite{Kilic2022}.

Here, we demonstrate the compressed sensing of polarized signals with low-refractive-index, self-assembled encoders. The unexpected transformations from linear to elliptical polarizations in the {self-assembled grating structures} have only recently been reported in our work \cite{Feng2022}. We draw an analogy to the compound-lens imaging system of insects, which is covered with self-assembled corneal random and ordered grating nanostructures \cite{Lee2016, Stavenga_2005, Blagodatski2015}. Although nanostructures in nature are associated with anti-reflective, hydrophobic, or self-cleaning functions \cite{Bernhard1962, Watson2004}, the optical encoding function of their nanostructures has not been explicitly shown. A survey of the potential modes for insect-inspired imaging and object tracking (colorimetric, polarimetric filtering, coherent optical preprocessing, and compressive sensing) has recently been written \cite{Vuong2023}. Whether insect corneal nanostructures have a physical encoding function or not, our approach with self-assembled encoders illustrates the potential. 

A key component to our encoders is that they scatter light into lattice modes that are sensitive to the {direction of the incoming wave and its} polarization ellipticity. Our work not only expands on our understanding of the physical encoding that is available with self-assembled nanostructures, but also outlines general insight for their design {with} compressed sensing.

\section{Results}
\subsection*{Polarization-modulated Diffraction from PANI-HS}
\begin{figure}[htb]
\centering \includegraphics[width=1\linewidth]{ 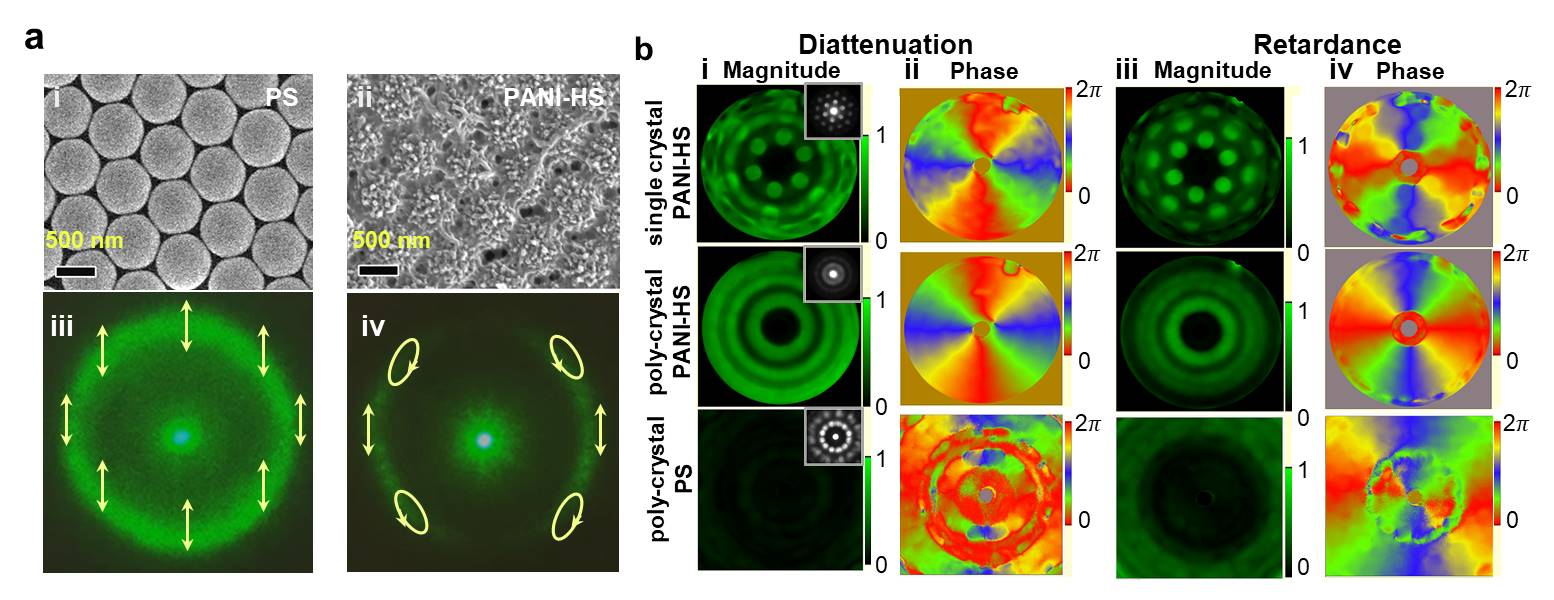} \caption{Polarization transformations of PANI-HS compared with those of polystyrene spheres (PS). (a) (i-ii) SEM images of PS and PANI-HS (720-nm lattice spacing) and corresponding (iii-iv) Debye or diffracted ring intensity patterns from vertical, linearly polarized (LP) light (532-nm wavelength). The measured diffracted polarization is overlaid in yellow. For PS, the diffracted intensity is uniform and the diffracted polarization remains vertical but for PANI-HS, both intensity and polarization vary spatially and the polarization ellipticity indicates spatially varying birefringence. (b) Conoscopy measurements showing (i-ii) diattenuation and (iii-iv) retardance magnitude and phase for single-crystalline PANI-HS, polycrystalline PANI-HS, and PS with 2-$\mu$m lattice spacing. Retardance and diattenuation are negligible in the PS samples compared to those of PANI-HS. } \label{Debye}
\end{figure}

We fabricate hierarchical 2D photonic-crystal films composed of {hydrochloric acid-doped} polyaniline nanofibers around hollow spheres (PANI-HS) based on an approach demonstrated in our prior work \cite{Feng2022}. Figures \ref{fig:intro2}(a-c) show the air-filled polymer grating structures. The film structure at the smallest level is composed of nanofibers surrounding a sub-micron-scaled spherical cavity, shown by the SEM image in Fig. \ref{fig:intro2}(b). We prepare PANI-HS films with two types of structures: polycrystalline (``ordered'') and short-range ordered (``random'') to investigate the impact of ordering on compressive sensing. Structural analysis was conducted using boundary counts and the 2D pair correlation function. 

{Figure \ref{fig:intro2}(d) shows the orientation of different grains, revealing boundary counts of 1.8 boundaries/$\mu$m and 3.3 boundaries/$\mu$m for the ordered and random samples, respectively.} In the boundary count method \cite{E112-13}, a grain is defined within a lattice with specific orientation; for our 6-fold symmetric assembly, a grain's orientation relative to an axis ranges from 0 to 60 degrees. A grain boundary occurs when neighboring domains deviate by a small angle. {Fig. \ref{fig:intro2}(e)} shows the 2D pair correlation function $g(r)$ \cite{Rengarajan_2005, Zhang_2012, Bohn2010} of the ordered and random samples alongside a perfect lattice. We also assessed ordering by comparing the width of the first-order peaks in the Fourier transform to that of a perfect lattice, yielding average orderings of 1.10 and 2.49 for ordered and random samples across three randomly selected areas. Additional detailed structural analysis is provided in Supporting Information Fabrication and Material Characterization of PANI-HS (Figs. S1 and S2). 

When light illuminates a larger, polycrystalline encoder area, we observe a diffracted ring also referred to as a Debye ring. The Debye ring and diffuse background patterns contain mixed, spatially multiplexed information, which arise from the mesostructure and the presence of nanofibers of the film. Figure \ref{Debye} illustrates the polarization-sensitive Debye ring from PANI-HS, which enables us to differentiate RCP, LCP, and unpolarized light with an LP camera. This polarization-modulated diffraction does not occur with traditional colloidal structures; while the diffracted light from polystyrene spheres (PS) colloids (and most solid colloidal structures) forms a uniform ring with unchanged polarization, the PANI-HS Debye ring intensity is broken and exhibits inhomogeneous polarization [Fig. \ref{Debye}(a)]. The hollow-sphere structure is important \cite{Retsch_2011, Baek_2020}: prior to the removal of PS, the PANI-coated PS films follow the behavior of PS rather than PANI-HS. PS alone exhibits higher intensities in the Debye ring compared to PANI but does not transform the diffracted light polarization significantly. The transformation from linear to elliptical polarizations is unexpected in {non-metal} colloidal grating structures. In fact, to the best of our knowledge, the polarization-modulated diffraction has only been recently reported in our work \cite{Feng2022}. {Approximately 70$\%$ of the light is transmitted and 10$\%$ of the transmitted power is diffracted into the PANI-HS Debye ring. The changes in polarization are sharper at the Debye ring for more ordered structures.} 
\begin{figure*}[h!]
  \centering
  \includegraphics[width=0.95\linewidth]{ 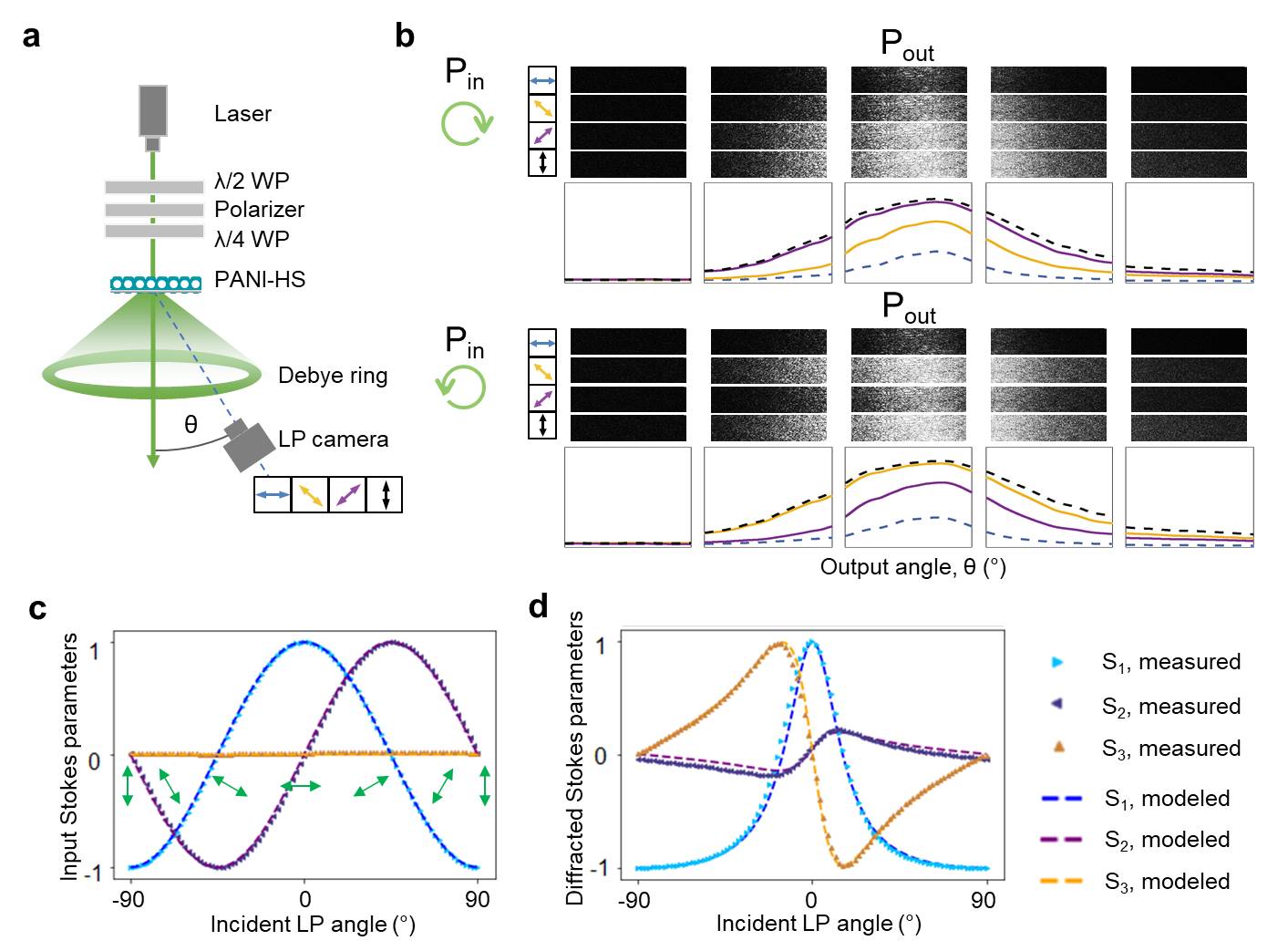}
  \caption{Polarization transformations with normally incident light. (a) Experimental setup where RCP and LCP light ($P_{\rm in}$) are produced with polarizer and waveplates (WP). Images at different polar angles $\theta$ are taken with a linear-polarization (LP) camera. (b) Linearly polarized images ($P_{\rm out}$) and vertical line traces over each polar angle (blue: 90LP, yellow: -45LP, purple: +45LP, and black: 0LP). Notice that the purple (+45 LP) and orange (-45 LP) traces are reversed for orthogonal input circular polarizations ($P_{\rm in}$). Modeled and measured Stokes parameters (c) for the input, where -90$^\circ$ (TE), 0$^\circ$ (TM), and 90$^\circ$ (TE) are the polarization angles with respect to the plane of measurement and (d) at the diffracted, Debye-ring polar angle ($\theta = 56^\circ$). 
}
  \label{fig:polexp}
\end{figure*}

To illustrate the polarization transformations of the film, we use Mueller matrix diffractometry, which shows the angle-dependent diattenuation (or polarization-sensitive losses) and retardance of the PANI-HS and PS arrays \cite{Arteaga_2017} [Fig. \ref{Debye}(b)]. In these measurements, the Mueller matrix polarimeter images the rear focal plane of the 50x (0.95 NA) objective to obtain complete polarization-resolved maps of the directional distribution of light. The complete Mueller matrix for each image in Fig. \ref{Debye}(b) is provided in the
Supporting Information Diffractometry Measurements (Fig. S3). {The high magnitudes of diattenuation and retardance in PANI-HS indicate that the material exhibits strong polarization-selective properties and significant birefringence, which is less pronounced in the PS film.} The comparison of measurements that probe single crystals vs. grain boundaries across multiple crystals [Figs. \ref{Debye}(b)i-ii] shows similar diattenuation and retardance {patterns}. This symmetry of the diattenuation and retardance allows us to rule out stress-induced birefringence in the coating \cite{Plekhanov_2011}. Also, since similar trends are observed for single and multiple crystals, the birefringence is not associated with the grain boundaries or defects in the self-assembled films \cite{GlushkoThesis}. 

{The PANI-HS response to incident polarization ellipticity is measurable with an LP camera or a camera with LP-filtered sensor pixels [Fig. \ref{fig:polexp}(a-b)].} To sense the handedness of the CP light, the key asymmetry arises in the tilted axes of the diffracted polarization; if this retardation is not present, then the CP component is not sensed. This effective birefringence is intriguing. The tilt of the linear polarization axis, also known as the azimuth of the polarization ellipse, depends on the CP handedness. The diffracted light appears as if it passes through a rotated retarder and diattenuator; when PANI-HS is illuminated with RCP and LCP, the Debye ring appears as a vector vortex ring, where the azimuth of the polarization ellipse is opposite for RCP and LCP. With an LP camera (Thorlabs CS505MUP1), four similar images with 0$^\circ$, 45$^\circ$, 90$^\circ$, and 135$^\circ$-angled LP filters are taken and summed vertically (along the axis perpendicular to the diffraction) [Fig. \ref{fig:polexp}(a)]. As shown in Fig. \ref{fig:polexp}(b), RCP and LCP are differentiated by the +45 and -45-LP components, which flip in power depending on the circular-polarization helicity. The relative strength between these two detected +45 and -45-LP signals reveals the relative phase between transverse electric (TE) and transverse magnetic (TM)-polarization components. This encoder with an LP camera achieves full-Stokes sensitivity.


The diffraction from nanofibrous PANI-HS films appears as the ``pooled'' scattered light from a single dipole scattering event with additional birefringence. A simple, coherent model fit to our measurements identifies scattering and retardance as the two underlying contributions to the polarization transformation. The polarization transformation is viewed as a rotated polarizer $\mathbf{J}_{\rm pol}$ and rotated waveplate $\mathbf{J}_{\rm wp}$,
\begin{eqnarray}
  \mathbf{E}_{\rm out} &=& \mathbf{J}_{\rm tot}\mathbf{E}_{\rm in}\\
	&=&[a\mathbf{J}_{\rm pol} + (1-a)\exp(i \gamma_{\rm rel}) \mathbf{J}_{\rm wp}] \mathbf{E}_{\rm in}, \label{mod}
\end{eqnarray}
where $a$ represents the relative weight and $\gamma_{\rm rel}$ represents the relative phase between $\mathbf{J}_{\rm pol}$ and $\mathbf{J}_{\rm wp}$. The rotation of $\mathbf{J}_{\rm pol}$ and $\mathbf{J}_{\rm wp}$ depend on the sampled azimuthal location on the Debye ring $\phi_0$:
\begin{eqnarray}
  			\mathbf{J}_{\rm pol} &=& \mathbf{R}(-\phi_0) 
  			\begin{bmatrix} 0&0\\0&1\end{bmatrix} \mathbf{R}(\phi_0) \\
      \mathbf{J}_{\rm wp} &=&\mathbf{R}(-\phi_0)
      \begin{bmatrix} 1&0\\0&\exp(i\gamma_{\rm ret})\end{bmatrix}\mathbf{R}(\phi_0), \label{mod2}   
\end{eqnarray}
where $\phi_0$ is measured in the sample plane with respect to the $x$-axis, and $\mathbf{R}(\phi_0) = \begin{bmatrix} \cos(\phi_0)&-\sin(\phi_0)\\ \sin(\phi_0)& \cos(\phi_0)\end{bmatrix}$
is the rotation matrix associated with azimuthal angle $\phi_0$, which is the LP polarization angle with respect to the incident plane. 

{At normal incidence, the fitting parameters to Eq. (\ref{mod}) are $a$ = 0.75, $\gamma_{\rm ret}$ = -70$^\circ$, and $\gamma_{\rm rel}$ = 85$^\circ$. These parameters show significant retardance. Figures \ref{fig:polexp}(c-d) show the input and transformed Stokes parameters generated by this model. The close agreement between our model and measurements indicates that the eigenmodes are linearly-polarized parallel and perpendicular to the $k$-vector at {azimuthal }angle $\phi_0$. These parameters indicate that the polarization modulation is distinctly measurable. The small deviations in ${\rm S}_2$ between our model and experiment may be attributed to misalignment of the incident angle of the sample or variations in the PANI surface flatness.}

\begin{figure*}[h!]
  \centering
  \includegraphics[width=.8\linewidth]{ 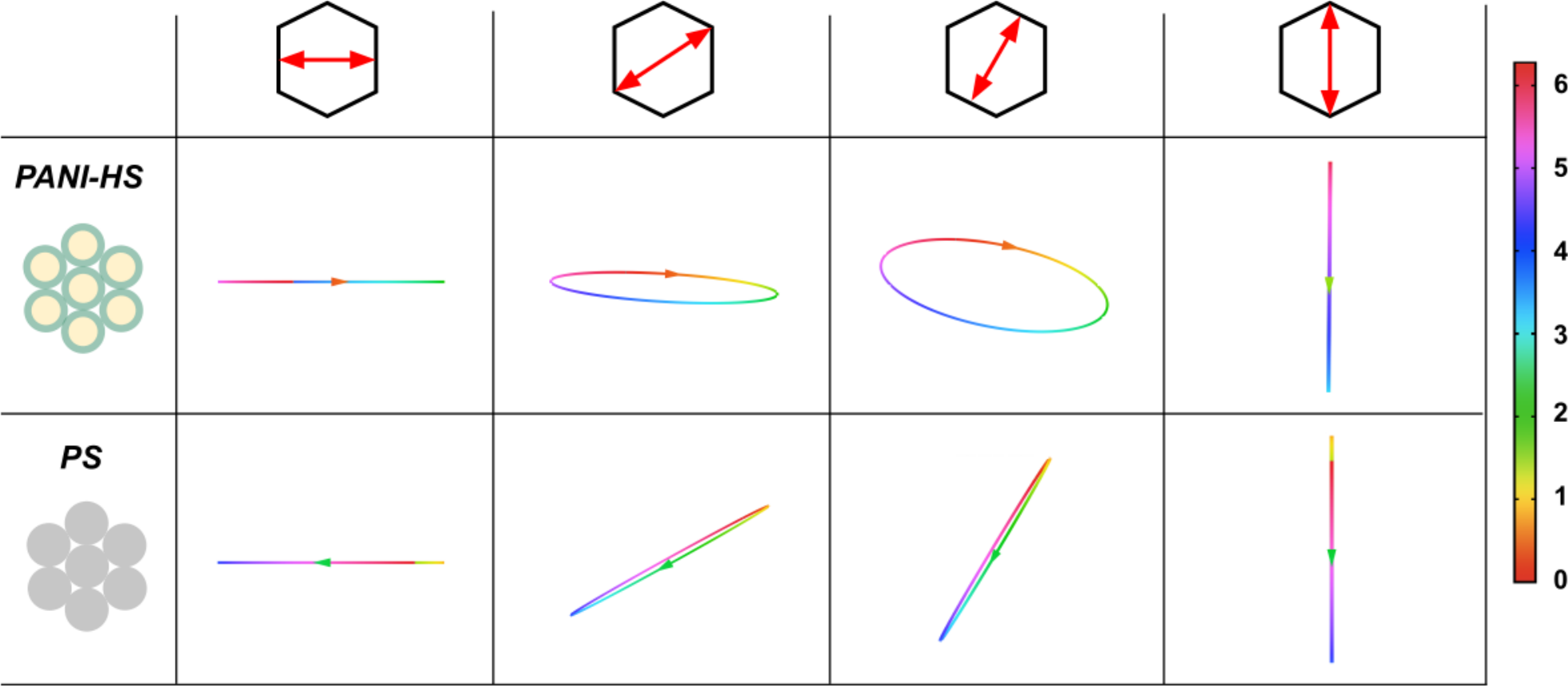}
  \caption{Numerically simulated Jones polarization plots of the diffracted mode by the PANI-HS (second row) and PS solid spheres (third row) arranged in a hexagonal lattice. The lattice orientation and incident linear polarization are shown by red arrows (first row). 
}
  \label{fig:COMSOL}
\end{figure*}


It can be stressed that the polarimetric sensing is possible because the azimuthal position in the Debye ring acts as the analog of a polarimeter-based rotated retarder with diattenuation in front of the detector. This is also what the Mueller matrix diffractometry images show {(See Supporting Information)}.

We measure that the degree of polarization {in the far field} is over 95\%; for our thin monolayer encoder, the assumption of polarization coherence is reasonable. {The full transformation from linear-to-circular polarization is observed with non-normal angles of incidence when light is composed of both TE and TM polarizations at an 8.5-degree angle of incidence (Supporting Information Fig. S4).} We observe circular-to-linear polarization conversion as well as linear-to-circular, and note that other polarizations are tuned with an angle of incidence in a manner analogous to metasurfaces that enable full-Stokes transformation and polarization tuning.

{To further validate the developed model (see Eqs. \ref{mod}-\ref{mod2}) and gain additional physical insights on the polarimetric transformation of PANI-HS, we perform full-wave numerical analyses. Using the commercial software Comsol Multiphysics, we simulate the PANI-HS and PS arranged in an ordered hexagonal lattice. Details on the numerical simulations are reported in the Methods section. Figure \ref{fig:COMSOLm} shows the polarization of the diffracted mode describing the Debye ring for four different orientations of the electric field of the incoming LP plane wave at normal incidence. When the electric field of the incident wave is parallel to the horizontal and vertical axis of the lattice (see row 1 and columns 2 and 5 in Fig. \ref{fig:COMSOLm}), neither PANI-HS nor PS structures transform the polarization and the diffracted mode remains LP, similar to the incident wave.}

On the other hand, for an incident LP wave with an electric field not directed along either the horizontal and vertical axes of the lattice (see row 1 and columns 3 and 4 in Figs. \ref{fig:COMSOLm}), the PANI-HS structure performs a polarization transformation. The PANI-HS diffracts an elliptically polarized mode, while the PS structure diffracts mostly LP light. These results show excellent agreement with the experimental results [Fig. \ref{Debye}(a)], confirming the ability of the PANI-HS to manipulate light polarization. They also show the accuracy of the developed model [Fig. \ref{fig:polexp}(c-d)], which predicts the linear-to-elliptical polarization conversion of the PANI-HS. When the direction of the electric field of the incident wave deviates from the horizontal axis, the amplitude of the Stokes parameter S$_3$, which indicates the strength of elliptically polarized light with S$_1$ is different than zero, increases and returns to zero for a vertically polarized incoming wave.


\subsection*{Metaphotonic Signatures for Compressed Sensing}

\begin{figure}[htb]
  \centering
  \includegraphics[width=0.7\linewidth]{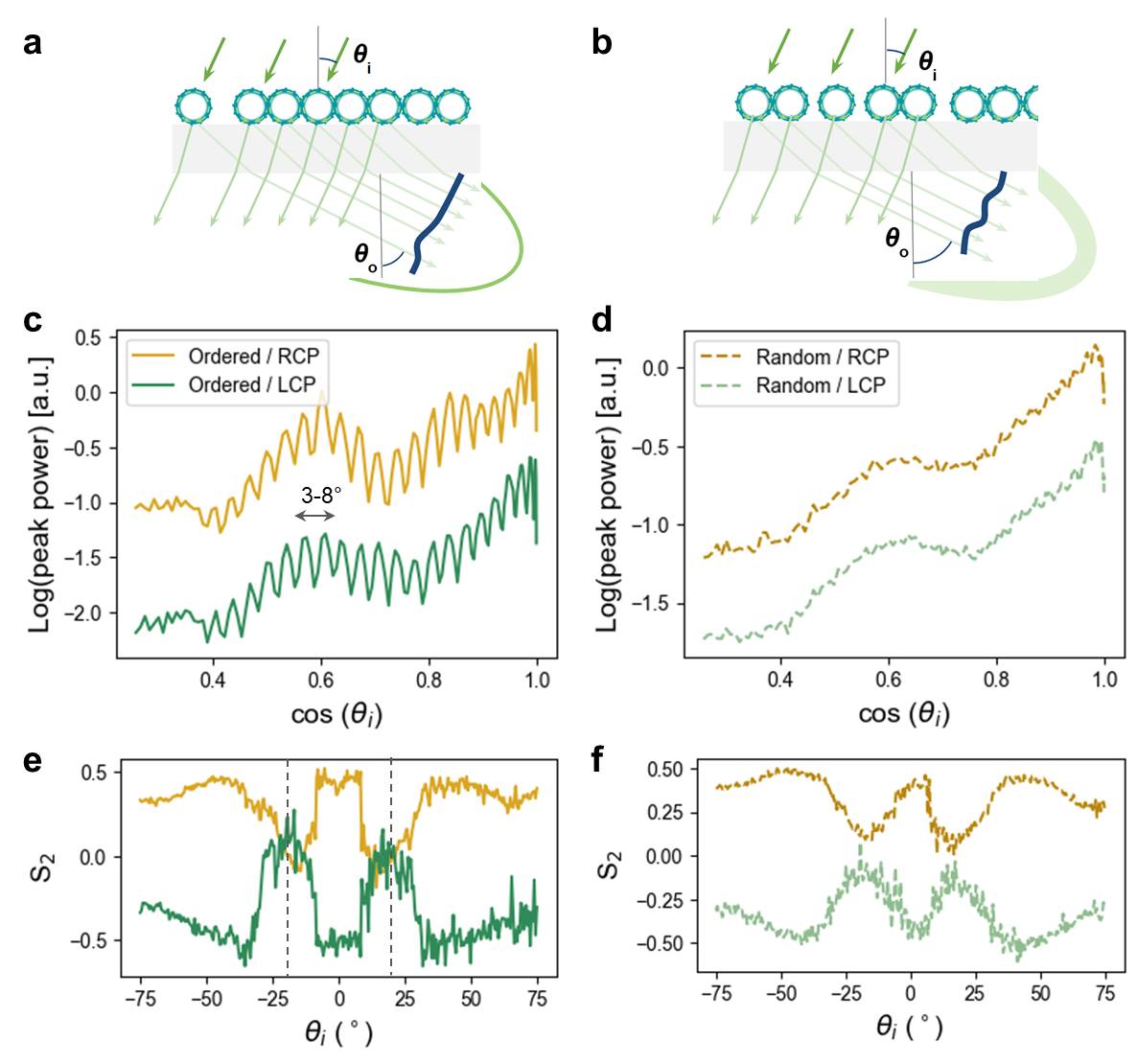}
  \caption{Physical encoding of ordered and random samples. Schematic of wavefront distortion from structure with small crystal grains for (a) ordered and (b) random hollow grating structures. The green arcs outline the first positive-order diffracted mode. The peak diffracted power along this ring for (c) ordered and (d) random hollow grating structure shows interference from the PANI-HS film index contrast with the glass substrate for RCP (top) and LCP (bottom). Modulated Stokes ${\rm S}_2$ parameter along the Debye ring for (e) ordered and (f) disordered structures.}
  \label{fig:NonnormalPower}
\end{figure}
{PANI-HS's diffraction-dependent polarization transformations are similar to the optical transformations generally associated with the nonlocal effects of metasurfaces or metagratings \cite{radi_metagratings_2017}. }Notably {however, an important aspect of the self-assembled encoder is a non-negligible degree of disorder, which offers an ``incoherence'' function for compressive sensing.} Compressed sensing is defined by two fundamental requirements: 1) {sparse} input information {(in the conventional imaging system [Fig. \ref{fig:intro1}(a)]), and} 2) ``incoherent'' output information \cite{Cand_s_2007} {(with the PANI-HS encoded system [Fig. \ref{fig:intro1}(b)])}. This second requirement is {not only influenced by the PANI-HS geometry but also} addressed by disorder in the sample \cite{Gigan2022}. 

We refer to the ``ordered''and ``random'' samples characterized in Fig. \ref{fig:intro2}. {The random PANI-HS samples offer a higher degree of incoherence compared to the ordered ones with broadened $k$-space angular modes, less-distinct fringes, and broadened modulation of the far-field polarization.} Figures \ref{fig:NonnormalPower}(a-b) illustrate the encoder wavefront distortion or incoherence function. When the sample is ordered (polycrystalline, with sharp pair-wise correlations and larger grains), then light is directed into angularly-sharp $k-$space diffraction peaks. When the sample is random and has short-range order, then the {diffracted interference} patterns are blurred. When the sample contains multiscale features such as nanofibers, the signal is spatially multiplexed into multiple $k-$space modes. The mode broadening and multiplexing enhances the capability for the single input-angle information to be captured in a few measurements. 

Experimentally, the output angle of the Debye ring $\theta_o$ follows a simple grating relation, however the peak power {\it values} at this diffracted angle varies approximately 2dB or 30\%. In Figs. \ref{fig:NonnormalPower}(c-d) we plot the peak transmitted power as a function of input angle captured for ordered and random structures. This reveals the interference patterns in the diffracted light patterns, which are more evident for ordered than random samples. The interference fringes are regularly spaced when we plot the peak intensity as a function of $k$ or $\cos(\theta_o)$. The fringes are easily interpreted as noise: one must sample the input angle $\theta_i$ with high resolution in order to observe the fringes clearly. These sharp interference fringes, which range in angular spacing $3-8^\circ$ from $\theta_i= -75$ to $+75^\circ$, encode beam pointing.

The diffracted polarization also varies as a function of input angle for ordered and random structures. This can be explained by the presence of weak lattice modes, which retard light and depend on the light's incident input angle and polarization. With simulations of ideal periodic lattices with effective medium theory (i.e., low refractive indices due to the air-filled fraction of the spheres), we attribute the polarization modulation to the presence of Rayleigh anomalies or TE-polarized lattice modes at $\theta_i=\pm 20^\circ$ (See Supporting Information). The refractive index data from prior PANI characterization \cite{Brasse_2019} is used. Experimentally, this angle of strong polarization modulation also coincides with increased absorption and diffraction. The ${\rm S}_2$ Stokes measurements or the tilt of the azimuth of the polarization ellipse, varies significantly between RCP and LCP inputs [Fig. \ref{fig:NonnormalPower} (e-f)]. To underline why the hollow-sphere structure is important: simulations do not show the presence of the Rayleigh anomaly with solid PS (with a higher refractive index) in similar lattice geometries. These results are consistent with prior experimental observations that lattices composed of PS spheres of similar dimensions do not significantly transform the diffracted light polarization like PANI-HS.  

When samples are ordered, the simulated resonances are narrow and located at an angle of incidence $\theta_i \approx \pm 15-20^\circ$. Strong TE modes are observed in ordered PANI-HS samples and disappear as samples are more random, evident from the specular transmission and reflection measurements with a lock-amplifier (See Supporting Information Measurements and Simulations, Fig. S7). {In contrast}, weak TM-polarized modes in the specular transmission and reflection vary in strength and occur at different incident angles across different experimental batch samples. These variations could relate to the {sensitivity of the TM modes to small variations in the sample, i.e., }grain mesostructure, fiber nucleation statistics, or PANI doping. However, over many batches and over years in ambient conditions, the sample TE lattice modes $\theta_i \approx \pm 15-20^\circ$ remain strong and consistent from batch to batch. The periodic modulations of ${\rm S}_2$ associated with the TE-polarized lattice modes occur with angular periods of $\sim 40^\circ$ and encode the incident state of polarization.

\begin{figure*}[h]
  \centering
  \includegraphics[width=0.9\linewidth]{ 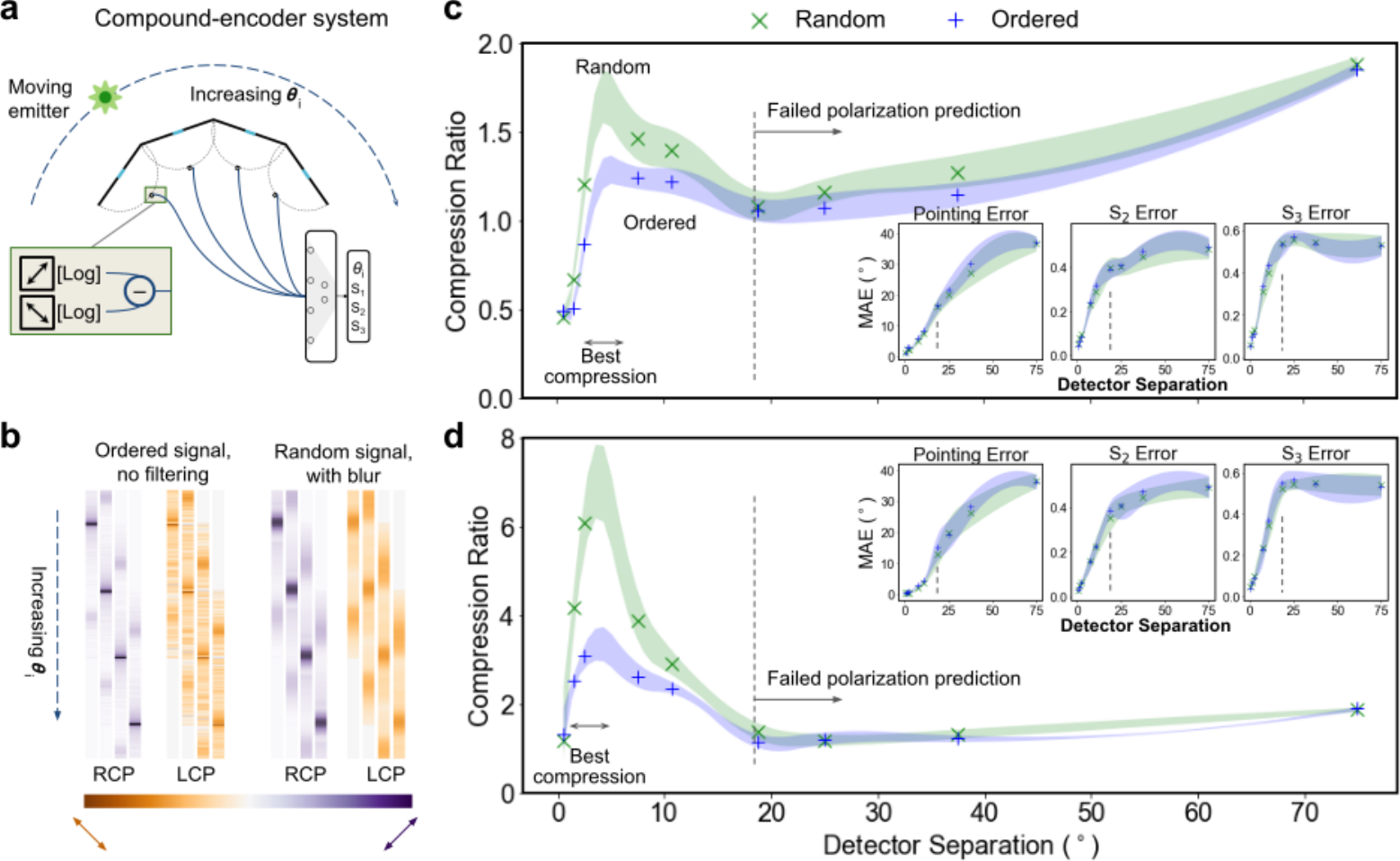}
  \caption{Multi-parameter estimation of incident beam polarization and pointing. (a) Illustration of 4-encoder-detector system for simulation with 150-degree field-of-view. The detector separation for this case is $\Delta \theta_o =37.5^\circ$. (b) Colormaps illustrating difference LP-signals given the geometry in (a) for raw experimental data of ordered sample (left) and 5-degree window-blurred sample (right). Compression ratios as a function of detector spacing with (c) raw data and (d) blurred data for ordered (blue) and random (green) samples. The markers show the mean values and shaded areas mark the upper and lower bounds from 15 different test simulations with polynomial fits between points. Inset: the pointing and polarization (${\rm S}_2, {\rm S}_3$) prediction errors. Dotted vertical lines mark Nyquist sampling of the pointing or when $\Delta \theta_o =$ Pointing MAE $\approx 20^\circ$. Best compression ratios are achieved with Nyquist sampling twice the frequency of the interference fringes or $\Delta \theta_o \approx 4^\circ$. }
  \label{fig:FilmPred}
\end{figure*}

 We simulate a nature-inspired compound, multi-faceted, PANI-coated system. Insects have compound encoded lenses with a sparse number of detectors, some of which sense linearly polarized light. They achieve a high-degree of visual acuity, which is often associated with micro-saccades or sharp head movements \cite{Srinivasan2004}. In an analogous manner, we collect the difference intensity signal from two orthogonal-LP detectors from each tilted encoder. We track polarized light from an emitter as it travels across a 150-degree field-of-view. For $m$ tilted encoders and $m$ detectors, the spacing between each detector 150$^\circ/m$ with the first and last detector located at $\theta_o = \pm 75^\circ \mp 150^\circ/2m$ [Fig. \ref{fig:FilmPred}(a)]. We make predictions of the beam pointing and polarization based on sparsely, equally-spaced sensors with a two-layer shallow neural network (SNN). More information about the model architecture is provided in the Supporting Information. 

To study the PANI-HS polarimetric encoding, we only consider input light emitters that are RCP, LCP, and $\pm45$LP, i.e., equal contributions of TE and TM polarizations associated with the ${\rm S}_2$ and ${\rm S}_3$ Stokes; to inform our understanding of the physical encoding, we initially exclude TE and TM inputs associated with the ${\rm S}_1$ Stokes to ensure the dataset is balanced. In other words, it is easy to differentiate TE and TM-polarized light by the varied reflection and refraction coefficients: if these inputs are in the dataset, the SNN learns to model their relations more readily and will not efficiently learn to differentiate the ${\rm S}_2$ and ${\rm S}_3$ polarization components.

For every input, the outputs are measured with a high-dynamic-range polarimeter (Thorlabs PAN5710IR1). We scan the output angle $\theta_o$ with 2.5-degree resolution and vary the incident angle $\theta_i$ in 0.5-degree increments for combinations of incident and output angles that span $\pm75^\circ$. The full dataset that shows full-Stokes characterization of the diffraction as a function of incident and output angle is illustrated in Sec. S3 of the Supporting Information. 

From this dataset that includes the full-Stokes characteristics at each angle, we consider the sparse ``detector measurements'' from two crossed LP detectors aligned $\pm 45^\circ$ from the sensing-plane. Figure \ref{fig:FilmPred}(b) shows colormaps of the LP-difference detected signals from each encoder given the geometry in Fig. \ref{fig:FilmPred}(a) for RCP and LCP. The LP-difference detected signals are the logarithms of the intensity signals multipled by the ${\rm S}_2$ measurements. The raw experimental data from ordered samples appear noisy and have narrow, angularly-sharp features; in fact, a significant amount of the information lies in the low-power measurements that are noisy. We smooth measurements with a Savitzky-Golay filter [Fig. \ref{fig:FilmPred}(b,right)]. This filter is analogous to the signal integration that is expected to be present in nocturnal insects and under low-light conditions \cite{Warrant2016, Warrant2017} or with micro-saccadic head movements. 

We define the compression ratio ${C_{\rm R}}$ as the improvement of the prediction error with encoding compared to the prediction error from conventional sampling \cite{Kilic2022}. The SNN predictions indicate two notable transitions, which are explained by the encoder characteristics. Figure \ref{fig:FilmPred}(c-d) shows the compression ratio ${C_{\rm R}}$ {where} the range of the error predictions $\varepsilon_{\rm Pointing}$, $\varepsilon_{\rm S_2}$, and $\varepsilon_{\rm S_3}$ {are} inset. When the detector spacing is larger than $20^\circ$, the SNN polarization prediction fails. Below a detector spacing of $20^\circ$, the model separates input ${\rm S}_2$ and ${\rm S}_3$ polarizations. When the detector spacing is $\sim 4^\circ$, we observe the highest values for ${C_{\rm R}}$. These transitions relate to the fringes produced by PANI-HS shown in Fig. \ref{fig:NonnormalPower}(c-f). 

In other words, the transitions in the SNN prediction errors are marked by the physical behavior of the film; the compression ratio is highest when the detectors sample the interferometric fringe and ${\rm S}_2$ polarization modulation at twice the angular frequency or {at} the Nyquist sampling criterion. The {role of the} encoder {is to map the} pointing and polarization {ellipticity} into angular-varying {intensity patterns that are, in this case, captured by crossed linear polarizers}. Regimes of undersampling and oversampling are identified above and below these transitions. The trends are similar across different SNN models and data preprocessing choices. 

Random PANI-HS samples achieve lower errors and higher values of ${C_{\rm R}}$ than ordered PANI-HS but also carry higher variances over different test runs. The high variances indicate that there is a significant degree of information in the sharp angular $k$-space modes that are easily interpreted as noise. This variance is illustrated by the shaded areas on each plot, which mark the range of predictions from 15 runs with different test splits. The markers denote the mean. The prediction error decreases significantly with signal blur or filtering and plateaus with a wider Savitzky–Golay window. 

The ideal degree of compression occurs for a relatively sparse separation of sensors (2-5$^\circ$) {over a wide field of view ($\pm 75^\circ$). Since the ideal sampling depends on the thin-film interference fringes, thicker encoders (larger-diameter hollow spheres or multiple-layer geometries) may be used to decrease the separation of sensors. It is expected, however, that such a change in the encoder will result in trade-offs, i.e., either a narrower field of view or more complicated back-end NNs.} Nevertheless, our results suggest that {with the compression ratios demonstrated here} one could achieve megapixel camera resolution with a kilopixel sensor array, particularly with further optimization with the NN architecture and the PANI-HS nanostructure.

\subsection*{Discussion}

The ability to encode full-Stokes polarization and achieve polarimetric compressive sensing is implemented with the back-end computing and the physical encoding from a self-assembled film. In the past, the encoder algorithmic function is programmed with digital arrays of micromirrors \cite{Chang2018, Wetzstein_2020}, pixelated spatial-light modulators \cite{Muminov2020, Wang2022NC}, and printed polarizer arrays or masks \cite{Li2022}. While these technologies are well established, most digitally-controlled technologies operate on the principles of filtering, reflection, or refraction, which alone cannot spatially separate RCP, LCP, and $\pm45$LP components (i.e., ${\rm S}_2$ and ${\rm S}_3$). 

To spatially separate beams with different polarization ellipticities, the interaction between polarization and pointing or an optical spin-orbit interaction is needed \cite{Vuong_2010,Bliokh2015}. Optical spin-orbit interactions are conventionally operationalized in metasurfaces with top-down fabrication paradigms and are therefore composed of periodic, high-refractive-index or metal nanostructures \cite{Xiong2023, Ling2015, Kwon2020,Zhou2020, Yang2018,Arbabi_2018, Ding_2021, Cheng_2020, Basiri_2019,Wang_2021,Balthasar_Mueller_2016}. Their ordered and periodic structures, while theoretically tractable {\i.e., straightforward to simulate and analyze with periodic boundary conditions}, do not provide the wavefront distortion and signal mixing necessary for compressed sensing. Notably, it is the ``imperfections'' associated with self assembly (i.e., randomly-oriented nanofibers and sample grain disorder) that provide the structure for physical encoding by which compressive sensing of pointing and polarization is achieved. Moreover and to a large extent, we observe that random structures achieve higher degrees of compression than ordered structures [Fig. \ref{fig:NonnormalPower}(a-b)]. 

 However, while random media provides the incoherence function for compressive sensing, too much disorder is difficult to decode. For example, the wave mixing of multimode fiber and multiple-scattering materials is enhanced in longer fiber or with longer optical path lengths in which the decoding algorithms are more involved or less accurate \cite{Borhani2018, Liutkus2014, Gigan2022, Asif_2017, Peng_2019, Antipa_2017, Guo2020, Pan2020}. In our work, random encoders also result in {less reliable predictions, i.e.,} a higher variance in their sensing performance [Figs. \ref{fig:FilmPred}(c-d)]. Even so, we show that the neural network learning transitions are predicted by the physical characteristics of the metaphotonic thin film. This means that we can, with physical understanding of a thin, self-assembled metaphotonic structure, design for optimal sampling and performance.

PANI-HS is a solution-processed metasurface \cite{radi_metagratings_2017} that spatially separates polarization components and whose thin-layer meso-order is easily decoded with a simple SNN \cite{Weng2023}. The self-assembled hollow-sphere diffractive films are also lightweight and low-cost, and their fabrication does not require expensive top-down techniques. The primary limitation of PANI is that it may degrade above temperatures of 140 ${^\circ}$C \cite{Kulkarni1989}, however other shell materials may be explored in lieu of PANI. 

\section{Conclusion}
In conclusion, we have demonstrated facile polarimetric compressed sensing with random self-assembled hollow-sphere encoders. An intermediate regime between order (lattices) and disorder (incoherence) is necessary to encode for polarimetric compressed sensing. In PANI-HS, the polarimetric compressed sensing is achieved by three primary responses: an effective birefringence via retardance that modulates the diffracted polarization via lattice effects; an effective thin-film interference that modulates the transmitted intensity patterns for pointing via refractive index contrast; and the broadening of the otherwise-sharp angular $k-$space diffracted features that provide incoherence or wavefront distortion via disorder or blur. Our physical encoder functions as a metasurface and a random scattering film, and offers fast and simple SNN decoding for rapid operations. Moreover, we have demonstrated that knowledge of the metaphotonic encoding (interference and polarization modulation) translates to best sampling for optimal performance.

Hollow encoders produced by templated colloidal assembly represent some of the simplest air-filled lattice designs that serve as a solution-processed metasurface. Our research on templated, hollow encoders not only offers insight about the vision systems of insects, but also frames a class of self-assembled materials for optical sensing and imaging. Our results with a solution-processed and mostly-air metaphotonic film open avenues for inexpensive, encoded large-area aperture and windows, for high-dimensional encoded computational cameras, including lightweight, high-speed, low-power, drone cameras and other polarization-sensitive edge computing applications. 

\section{Methods}

\subsubsection*{Chemicals}
Styrene, potassium persulfate (KPS), sodium dodecyl sulfate (SDS), aniline, ammonium persulfate (APS), were purchased from Sigma-Aldrich. Toluene, HCl, $H_2SO_4$ and $H_2O_2$ of chemical reagent grade were purchased from Fisher Chemicals. All the chemicals were used as received without further purification.

\subsubsection*{Synthesis of PS nanospheres}
Monodispersed 720 nm PS dispersions were prepared by emulsion polymerization.\cite{Lu2011} 1 0.092 g SDS and 0.0836 g KPS were dissolved in 40 mL ethanol and 16 mL $H_2O$ in a 250 ml three-neck flask with $N_2$ gas. Then 3.76 mL of styrene was added under rapid stirring. The emulsion solution was heated to 70 °C and maintained for 8 h. The PS nanospheres were washed with ethanol 5 times before the assembly.

\subsubsection*{Interfacial assembly of PS nanospheres}
A 300 mL crystallizing dish was filled with water to provide the air-water interface for the assembly. A dispersion of 5 wt\% PS nanoparticles in a mixture of ethanol/$H_2O$ (v/v=1/1) was pumped at a speed of 0.01 mL/min to the air-water interface. The PS nanosphere assembly was collected on a glass substrate which has been pretreated with $H_2SO_4$/$H_2O_2$ (v/v=3/1) at 80 °C for 30 minutes. The PS nanosphere assembly was dried at room temperature and heated at 60 °C for 1 hour to help stabilize.

\subsubsection*{Synthesis of PANI hollow sphere (PANI-HS) film}
In a typical synthesis, 18 mL of concentrated HCl was added to 180 mL of $H_2O$, followed by the addition of 111 $\mu$L of aniline and 0.306 g of APS. The solution was stirred for 3 minutes to obtain a homogeneous PANI growth solution. The glass substrate with the PS assembly was then inserted into the PANI growth solution with the PS side down. After 15 hours, the PS-PANI film was removed from the growth solution. The sample was dried at room temperature, followed by removal of the PS nanospheres by toluene.

\subsubsection*{Characterizations}
The structure of the PANI fibers, the PANI-HS and the PS array was characterized by a Thermo Fisher Scientific NNS450 scanning electron microscope. The samples on glass substrates were sputter-coated with Pd/Pt and characterized at 20 kV. Digital images of the Debye rings were taken by using an iPhone camera. The polarization data was taken with a PAN5710VIS polarimeter. The digital images in Fig. \ref{Debye} were taken with a Thorlabs CS505MUP - Kiralux polarization camera.

{
\subsubsection*{COMSOL numerical simulations}
The numerical simulations were performed using the commercial software COMSOL Multiphysics. The refractive index used for PANI-HS were tabulated from digitized images in previously-published ellipsometry measurements of acidic PANI \cite{Brasse_2019}. Both PS solid spheres and PANI-HS were modeled as infinite structures through a hexagonal unit cell with periodic boundary conditions. Two ports located at a distance five times the side of the unit cell were used to illuminate the structures and absorb the diffracted waves. Finally, a triangular mesh was implemented with minimum and maximum sizes of 0.01$\lambda$ and 0.1$\lambda$, respectively, to ensure accurate results.}

\subsection*{Data Archival}
The dataset and code to process the results are available.

\begin{acknowledgement}
Authors gratefully acknowledge funding from DARPA YFA 19AP00036 and NSF DMR 1921034, which supported all experiments and analyses with exception of the COMSOL simulations and Mueller matrix diffractometry. OA acknowledges TED2021-129639B-I00, CNS2022-136051 (Ministerio de Ciencia, Innovación y Universidades). S. Weiss fabricated the samples sent to OA. 
\end{acknowledgement}

\begin{suppinfo}

\section{Fabrication and Material Characterization of PANI-HS}
\subsection{Pair correlation}
\begin{figure*}[htb]
    \centering
    \includegraphics[width=.7\linewidth]{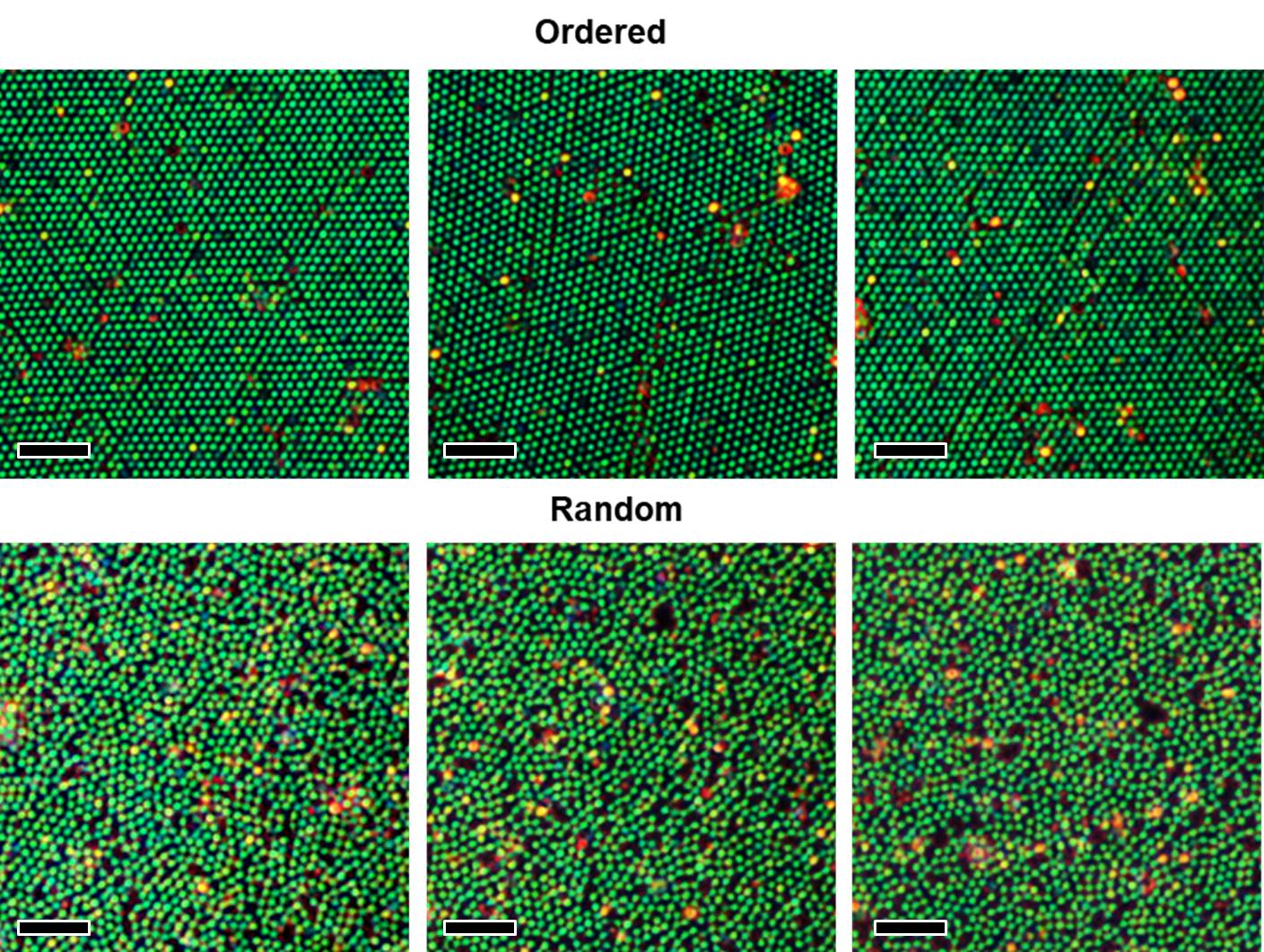}
    \caption{Optical images used to calculate the order parameters. All scale bars represent 5 ${\mu}m$. }
\end{figure*}
Previously we produced by a similar method polyaniline (PANI) inverse opal structures \cite{Feng2022}, where we identify the off-resonance dipole scattering from PANI nanofibers. While many optical applications leverage the strong localization of fields through nanophotonic resonances, off-resonant scattering confers opportunities to manipulate the polarization of light without tremendous concern for the phase changes induced by a single scattering event. By employing off-resonant scatterers with smaller-absorption cross-sections in periodic nanostructures, we ease the fabrication tolerances at the nanoscale and promote a higher degree of resilience with fabrication. In order to improve cavity effects within the domain of self-assembled structures, we fabricate hollow-sphere photonic-crystal films. The hollow-sphere structure doubles the volume of PANI compared to an inverse opal film \cite{Feng2022}; correspondingly, we observe an increase in the strength of the scattered, diffracted light.

 Light incident on an ideal 2D lattice diffracts largely into a hexagonal array of points that correspond to the six reciprocal lattice vectors of the particle array. However, if the lattice exhibits some degree of rotational disorder, the diffraction will form a continuous ring that is often referred to as a Debye ring, which refers to light diffraction into a specific angle $\theta_d$:
\begin{equation}
    \theta_d = \sin^{-1} \left(\frac{2\lambda}{\sqrt{3}d}\right),
\end{equation}
where $\lambda$ is the wavelength of the incident light, $d$ is the diameter of the sphere. This angle of diffraction also represents the circle in k-space with radius $k_\perp$ of the grating diffracted mode,
\begin{equation}
    k_\perp = \frac{2\pi}{\lambda} \sin(\theta_d) =  \frac{4\pi}{\sqrt{3} d}
\end{equation}
To further quantify the degree of ordering by using this synthesis approach, we analyze the distribution of PANI-HS in three randomly chosen areas, and calculate their 2D pair correlation function, $g(r)$ \cite{Rengarajan_2005, Zhang_2012}, 
\begin{equation}
g(r) = \frac{1}{\langle \rho \rangle}\frac{dn (r, r+dr)}{da(r, r+dr)}
\end{equation}	
where a is the shell area, $dn$ is the number of hollow nanospheres that lie within a shell, $r$ is the distance from an arbitrary origin, and $dr$ is the shell thickness. The statistical average of hollow nanospheres is normalized by the average particle number density $\left< \rho \right>$  and the sampling area $da = 2\pi rdr$. We analyzed optical images of 900*900 pixels (Figure S1.1) with $r$ = 11 and $dr$ = 0.725. The pair correlation function of PANI-HS and a perfect lattice is plotted in Figure 2a in the main text. 

Figure 2e in the main text also shows the first peak of the Fourier transform of each $g(r)$. We compare the full width at half maximum (FWHM), $\kappa$, for the first peak of $g(r)-1$ to that of a perfect lattice ($\kappa_0$). The ratio of $\kappa/\kappa_0$ is used to quantitatively determine the ordering of the photonic structure, and average from the three areas. A structure with $\kappa/\kappa_0 \leq 1.5$ is considered very highly ordered. The averaged $\kappa/\kappa_0$  for the ordered and random PANI-HS are 1.10 and 2.49 with standard deviation of 0.07 and 0.26.

\subsection{Mesoscale Grain Size Determination}
\begin{figure}[ht]
    \centering
    \includegraphics[width=\linewidth]{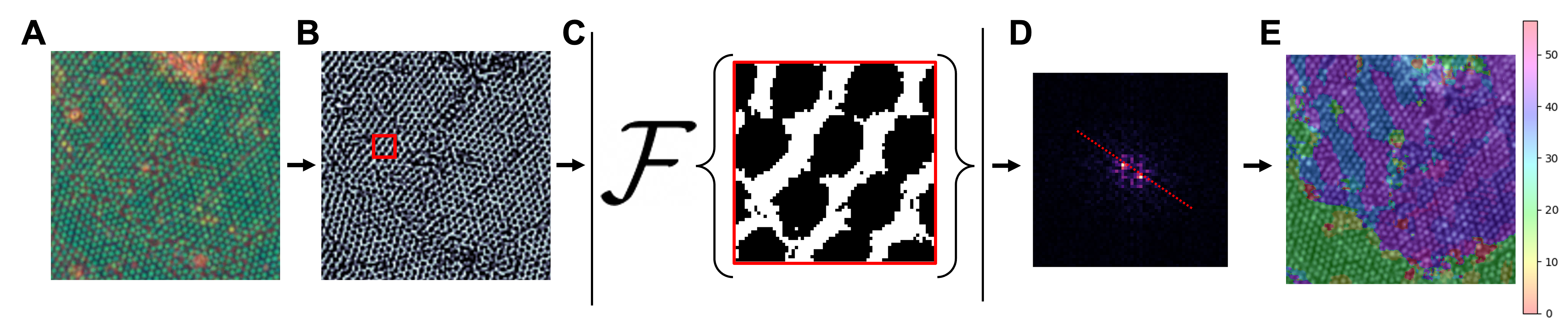}
    \caption{(A) Cropped portion of the ordered PANI sample. (B) Same image as `A' but with Guassian Adaptive thresholding applied. (C) The absolute value of the 2-D Fourier transform of the small window shown in (B). (D) The result of the operation illustrated in `C'. The direction of the grating is defined as the line that joins the first-order diffraction spikes (indicated by a red dashed line). (E) The windowing process shown in `C' and `D' is scanned across the image `B'. The resulting angles are perpendicular to the line in `D'. The grain angle mapping is overlaid on `B'.}
    \label{fig:grain determination}
\end{figure}

The process for determining local grain orientation is illustrated in Fig. \ref{fig:grain determination}. The process consists of an analysis of the angle of the first order spikes that appear in the absolute value of the Fourier transform of any grating structure. 
First, a high contrast image of the film must be made. We do this with Gaussian Adaptive Thresholding implemented through the OpenCV python library. A $64\times64$ window is made in the high contrast image. We take the 2-D Fourier transform of this window and determine the orientation of the first order peaks. The angle of the two peaks is determined with respect to a horizontal line. This denotes the normal of the actual grain direction. If no distinct first-order peaks appear, then the window is skipped and the angle determination is determined by its nearest neighbor. 
The orientation of the grain is attributed to the pixel at the center of the window. Since the grating structure has 6-fold symmetry, all grain angles, ($\phi_i$, $\phi_j$), that satisfy 
\begin{equation}
    \phi_i\bmod 60^o=\phi_j\bmod60^o
\end{equation}
implies that $\phi_i = \phi_j$. This is evident in Fig. \ref{fig:grain determination}(E), the grain map.

To determine Linear Intercerpt Density (LID) \cite{E112-13}, 10000 chords are constructed via the choice of two random, not identical, points in the grain map. These chords are rasterized into a number of pixels, $p_i$, along the $i^{th}$ chord. One of the ends of the chord is chosen as the start; from there, the grain orientation at that pixel is noted and the grain orientation of the next pixel is read. If the grain orientation of the next pixel is sufficiently different than the orientation of the previous pixel ($>\pm.5^o$), then $1$ is added to the boundary counter, $b_i$. The average grain density is determined by
\begin{equation}
    LID = \frac{1}{N}\sum^N_{i=0}\frac{b_i}{p_i}
\end{equation}
This process was calculated on 64 images of the ordered sample and 6 images of the disordered sample. The resultant LIDs were determined to be $1.7 \pm 0.5$ boundaries/$\mu$m and $3.1 \pm 0.4$ boundaries/$\mu$m for the ordered and disordered samples respectively.

\clearpage

\section{Polarization Characterization of PANI-HS}
\subsection{Diffractometry measurements}\label{sec:conoscopy}
Mueller matrix diffractometry measurements are performed using Mueller matrix microscope equipped with a Bertrand lens. The sample is illuminated with a collimated beam, which is collected by a high NA microscope objective. The Bertrand lens focuses the back focal plane of the objective onto  the camera, so that  it is possible to image the angle-resolved diffraction of the colloidal samples. This instrument uses a computer controlled polarization state generator and a polarization state analyzer to probe the sample with different polarization states, allowing the automatized measurement of the 16 Mueller matrix elements. Further details of the instrument used are available in a prior publication \cite{Arteaga_2017}.

The Mueller matrix analysis of the single-crystal PANI-HS, poly-crystal PANI-HS, and polycrystal PS with 2-$\mu$m spacing is shown in Fig. \ref{fig:Conoscopy}. In these measurements the light source was a 550 nm LED and the objetive used is a Planachromat 50× with 0.95 NA. In the case of single crystals (Fig. S2A) the observed diffraction pattern is a direct representation of the reciprocal lattice of crystal. 

\begin{figure*}[ht]
    \centering
    \includegraphics[width=\linewidth]{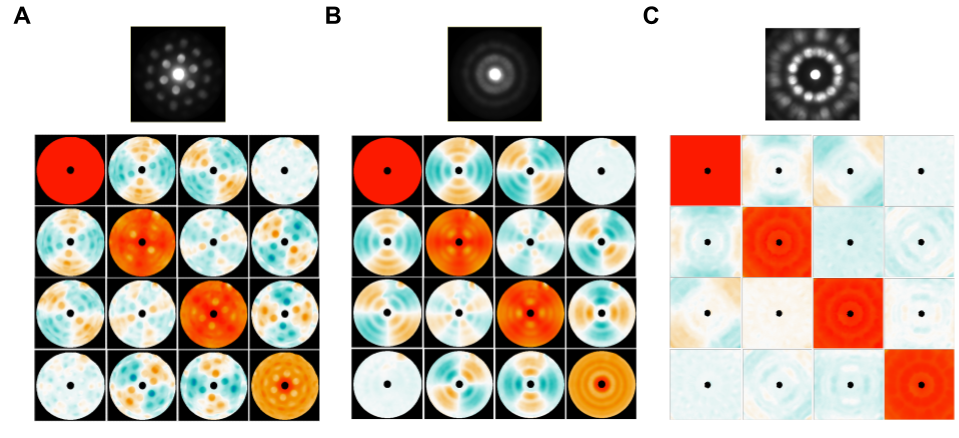}
    \caption{Mueller matrix diffractometry results on (A) a single-crystal (PANI-HS) , (B) poly-crystal (PANI-HS)  and (C)polycrystal (PS). All Mueller matrix elements have been normalized to the first element. Top images show the measured intensity patterns. 
}
    \label{fig:Conoscopy}
\end{figure*}
\clearpage

\subsection{Conversion of Jones to Stokes polarization measures}
We convert our model from Jones to Stokes vectors since each representation has different advantages. The input and diffracted Jones vectors $\mathbf{E}_{in} = E_o[A_x, A_y \exp(-i\delta)]^{T}$ are shown with Stokes measures ${\rm S}_1 = A_x^2-A_y^2$, ${\rm S}_2 =2A_xA_y\cos(\delta)$, and ${\rm S}_3 = 2A_xA_y\sin(\delta)$, where $A_x^2 + A_y^2 = 1$.  Additionally, $\delta = \pm \pi/2$ and ${\rm S}_3 \pm 1$ for RCP/LCP, respectively.  

\subsection{Linear-to-Circular Polarization Conversion}
The full transformation from linear-to-circular polarization is observed with non-normal angles of incidence when light is composed of both TE and TM polarizations at an 8.5-degree angle of incidence. With the ordering ratio of $\kappa/\kappa_0 \leq 1.5$, we observe circular-to-linear polarization conversion as well as linear-to-circular, and note that other polarizations are tuned with an angle of incidence in a manner analogous to metasurfaces that enable full-Stokes transformation and polarization tuning [See Fig. \ref{fig:lp2cp}].

\begin{figure*}[ht]
    \centering
    \includegraphics[width=.7\linewidth]{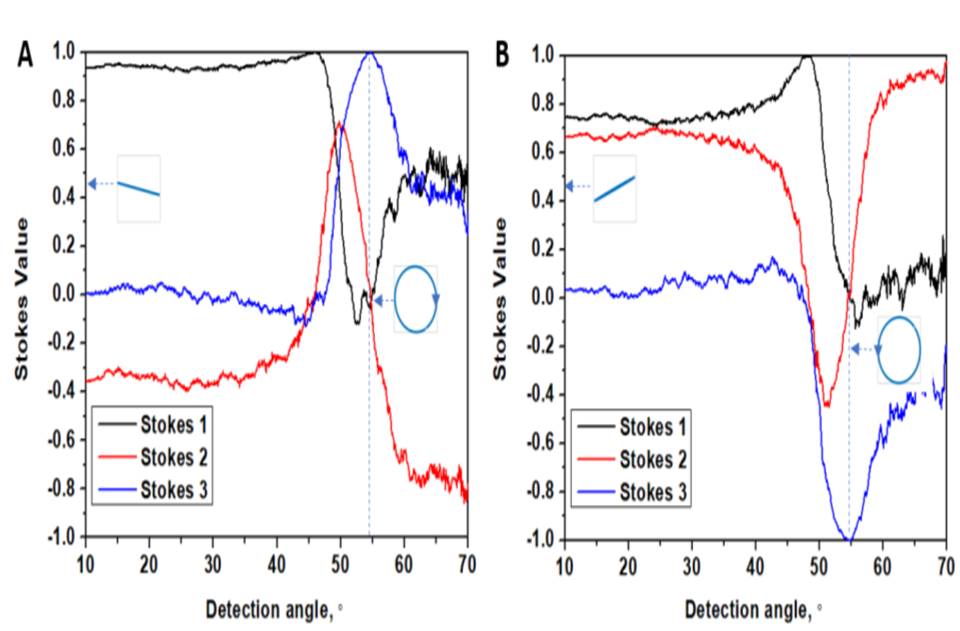}
    \caption{Conversion from linear to circular polarization. The polarization conversion is  $-10^\circ$ LP to RCP for (A), and $+20^\circ$ LP to LCP (B). The incident angle is $8.5^\circ$ for both measurements.}
    \label{fig:lp2cp}
\end{figure*}
\clearpage

\section{Polarization Dataset}
\subsection{The dataset}

We create a dataset where we interrogate the sample with $+45^\circ$, $-45^\circ$ linearly-polarized ($\pm45$LP) and right and left circularly-polarized (RCP and LCP) light and measure the power and polarization for different incident angles (-75 to 75$^\circ$) and output angles (-75 to 75$^\circ$ with respect to the normal of the sample). The input angles vary $0.5^\circ$ and the output angles are measured in $2.5^\circ$ increments. 

\begin{figure}[bht]
    \centering
    \includegraphics[width=\textwidth]{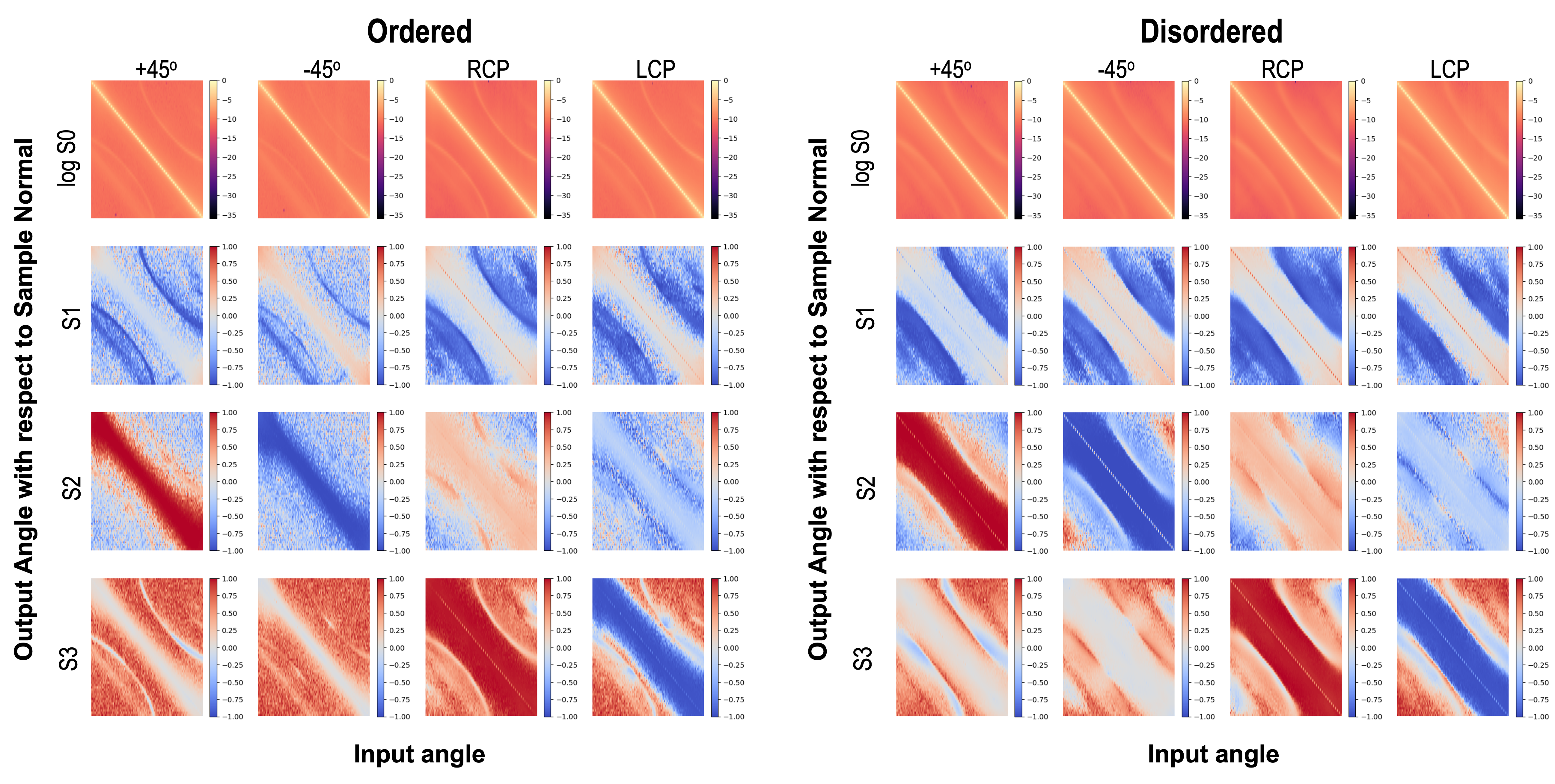}
    \caption{Stokes measurements of ordered (left) and disordered samples for input light of $\pm 45^\circ$, $RCP$, and $LCP$ polarizations. Both input angle and output angle (with respect to sample normal) vary between $-75^\circ$ and $75^\circ$.}
    \label{fig:low_intensity_S2S3}
\end{figure}

We collected data at higher intensity for the disordered sample with input light that characterizes a full Stokes input.

\begin{figure}[bht]
    \centering
    \includegraphics[width=.8\textwidth]{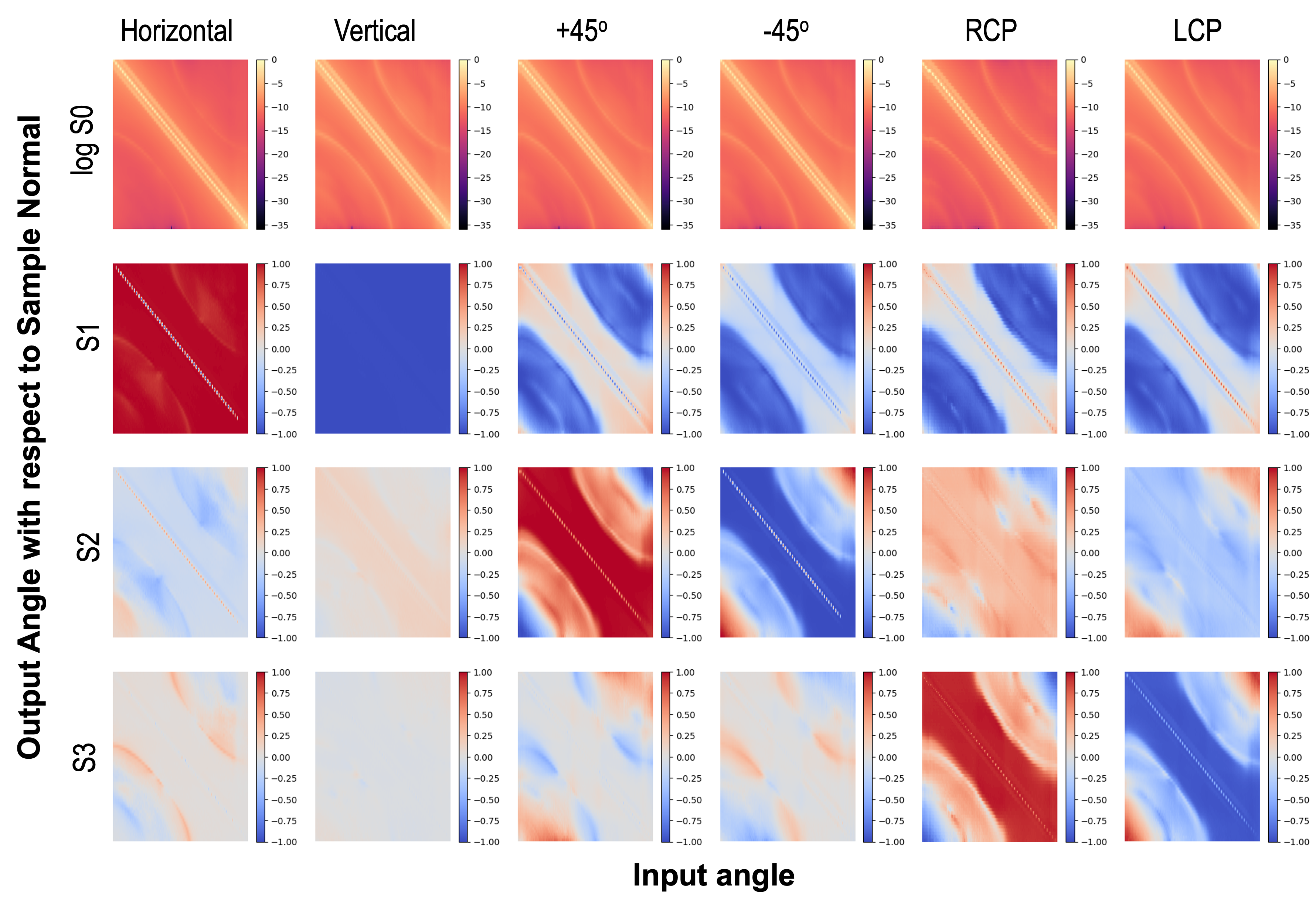}
    \caption{Stokes measurements of disordered sample at higher power for input light of horizontal, vertical, $\pm 45^\circ$, ${\rm RCP}$, and ${\rm LCP}$ polarizations. Both input angle and output angle (with respect to sample normal) vary between $-75^\circ$ and $75^\circ$.}
    \label{fig:enter-label}
\end{figure}


\clearpage

\section{Measurements and Simulations}

\subsection{Experiments with lock-in}
Fig. \ref{fig:COMSOL} shows further experiments and numerical simulations illustrating TE and TM transmission resonances. The transmission experimental setup uses a chopper and lock-in amplifier [Fig. \ref{fig:COMSOL}(A)i]. Ordered samples show the greatest degree of polarization modulation. TE-polarized fields exhibit a strong mode at 20-degrees angle of incidence, while TM-polarized fields exhibit weak dips at 5, 30, and 45$^\circ$ [Fig. \ref{fig:COMSOL}(A)ii]. The TE-polarized mode is prominent and consistent. 

\subsection{COMSOL}
We calculate the transmitted fields in a hexagonal lattice with COMSOL [Fig. \ref{fig:COMSOL}(B)i]. The refractive index data from prior PANI characterization \cite{Brasse_2019} is used. For incident TE and TM polarized light, we vary the azimuthal planes of incidence ($\phi = 0, 7.5, 15, 22.5$, and $30^{\circ}$), the polar angles of incidence  ($\theta = 0 .. 60^{\circ}$), and the different shell thicknesses $t= 40-120 $nm. A coherent weighting coefficient for each thickness, $c(t)$, is fit to minimize mean-squared-error to the experimental TE transmission curve shown in Fig. \ref{fig:COMSOL}(B)ii. We subsequently fit the numerically-weighted and coherently superposed TE and TM transmitted curves for this TE-fit $c(t)$, which shows good agreement with the experimental results. This indicates that the TE and TM dips in polarized transmission can, to some extent, be explained by a thin, solid shell lattice with a distribution around $t=100$-nm thickness. 

\begin{figure}[htb!]
    \centering
    \includegraphics[width=.8\linewidth]{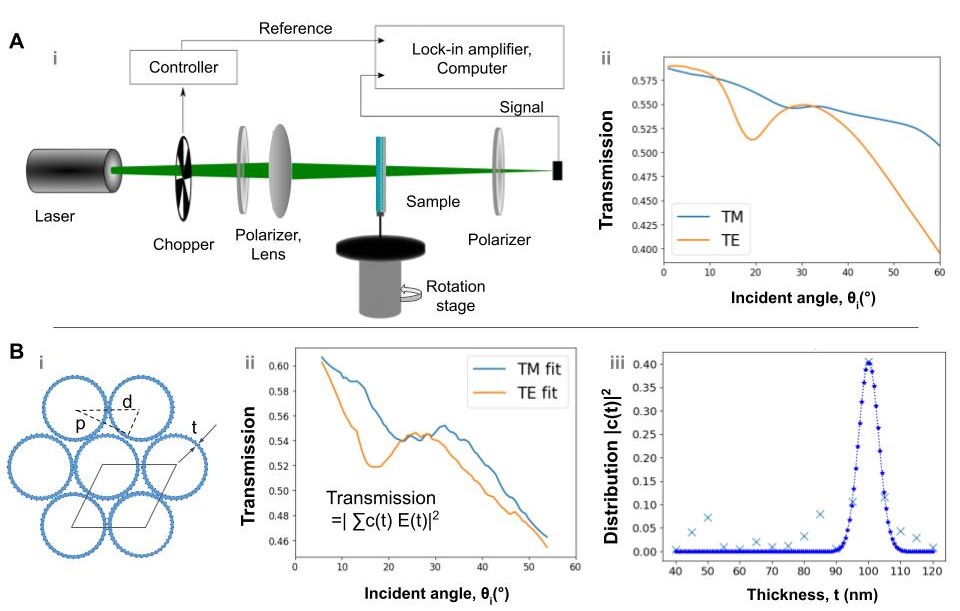}
    \caption{Polarization-dependent modes from hollow-shell photonic crystals. (a) Transmission measurements showing strong TE and weaker TM modes. (i) Experimental setup with referenced chopper and lock-in amplifier where the polarizers before sample and detector are both aligned for TM or TE transmission. (ii) Average measurement as a function of sample incident angle for incident TE and TM polarized light over 5 areas on a sample.  Shaded areas show maximum and minimum curves. (b) Numerically simulated transmission for a smooth and ideal hexagonal lattice of thin shells with PANI refractive index. (i) Shell lattice geometry for lattice spacing $p$ = $\frac{\sqrt{3}d}{2}=625$ nm and thickness $t$, which we vary from 40 to 120 nm. The solid line represents the unit simulation cell. (ii) Fit transmission weighted by a thickness-dependent complex coefficient $c(t)$. (iii) Fit probabilities $|c(t)|^2$ as a function of thickness $t$. A Gaussian-distribution fit to $c(t)$ is centered around 100 nm, which would agree with the distribution observed in SEM images.  
}
    \label{fig:COMSOLm}
\end{figure}

\subsection{Mie theory}

The optical properties of homogeneous and core-shell nanospheres, such as absorption, scattering, extinction efficiencies, and scattering patterns, can be determined using Mie's theory \cite{Chapter4,Chapter8}. The electromagnetic simulations of periodic/random arrays of core-shell spheres are a challenging task; for this reason, in some cases, it is desirable to model the core-shell structure as an equivalent sphere with an effective refractive index for simply the problem. 

With Mie theory, it can be shown that individual hollow spheres exhibit 20-times reduced specular reflection and 2-3 times transmission compared to individual solid PS spheres of the same size. This would explain why the hollow-sphere geometry is important for observing the polarization-dependent transformations: a greater percentage of light is back-scattered into nonlocal modes in the hollow-sphere lattice than a PS-sphere lattice. Results from Mie theory are shown in Fig. \ref{fig:Mie}.

\begin{figure}[htb!]
    \centering
    \includegraphics[width=.8\linewidth]{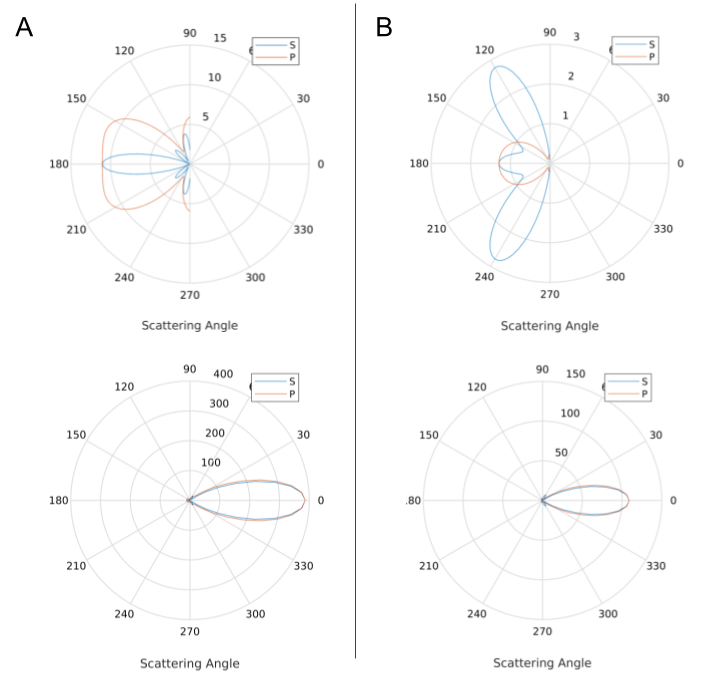}
    \caption{Mie scattering of solid (A) 720-nm diameter PS spheres (refractive index $n= 1.55$) and (B) 720-nm diameter, 60-nm thick shells of PANI (refractive index $n=2.2+0.1i$). Top images show back scattering and bottom images show forward scattering for TE (s) and TM (p)-polarized light.
}
    \label{fig:Mie}
\end{figure}

\subsection{Effective medium theory}

Effective medium theories provide a means to calculate the dielectric function of an equivalent sphere with the same radius and electromagnetic characteristics as a spherical core shell. Among various theories, we use the weighted average approach \cite{Lee2006} to compute the effective dielectric function,
\begin{equation}
    \varepsilon_{eff} = f\varepsilon_{c} + (1-f)\varepsilon_{s}.
\end{equation} In the given equation, $f = R_c^3 / R_s^3$ represents the volume fraction, where $R_c$ and $R_s$ are the radii of the core and shell, respectively. Additionally, $\varepsilon_c$ and $\varepsilon_s$ denote the dielectric functions for the core and shell, respectively. This yields an effective refractive index of $n=1.26 + 0.025i $. 

We simulate this 2-D photonic crystal lattice of spheres with these refractive indices with the S$^4$ package \cite{Liu20122233} and calculate the specular transmission and reflection of the TE (s) and TM (p) polarizations. The experimental transmission and reflective specular measurements are shown in Fig. \ref{fig:S4} for one orientation and for an average of seven, equally-spaced rotated orientations. With one orientation, we observe what appears to be a Rayleigh anomaly at an incident angle of 15 degrees for the TE polarization. With seven averaged orientations, the feature is broadened and extends to an angle of incidence of 20 degrees. We believe this lattice effect is a key feature in the PANI-HS enabling its polarization transformations.

\begin{figure}[htb!]
    \centering
    \includegraphics[width=.8\linewidth]{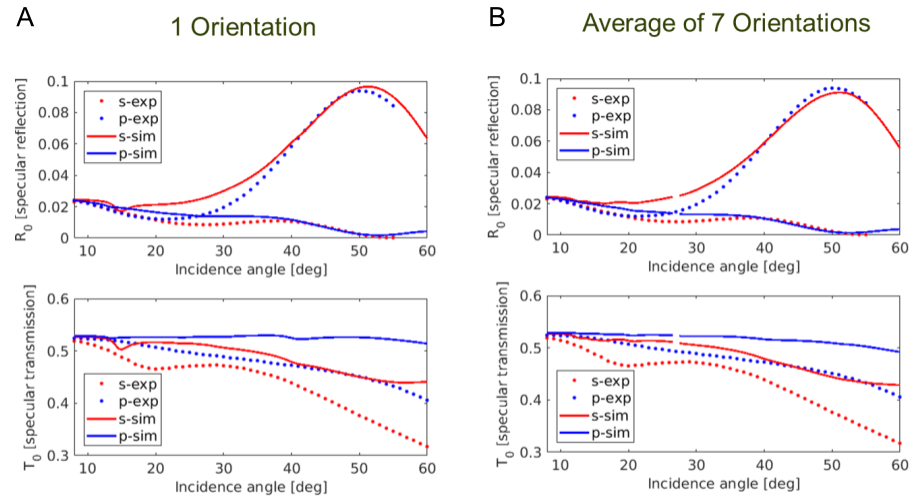}
    \caption{Specular reflection (top) and transmission (bottom) for (A) a single orientation and (B) average of seven rotated orientations for simulated 720-nm diameter, single-layer photonic-crystal lattice. (refractive index $n=1.26 + 0.025i $). The solid and dotted lines indicated simulated and experimental curves, respectively.
}
    \label{fig:S4}
\end{figure}

\section{Multi-parameter estimation with shallow neural networks (SNN)}
\subsection{Two-layer simple model}
We use a 2-layer dense shallow neural network (SNN) model in python tensorflow with multiparameter estimation to predict the beam direction and polarization. The system converges consistently with different degrees of variance depending on the test split. One reason our system is robust is that there is a minimal degree of nonlinearity; nonlinearity arises in the intensity modulus sensor measurements and in the SNN dual-layer activation functions. 

The SNN inputs are the stacked signals from $m$ detectors equally spaced 5-30$^\circ$ apart between $-\theta_{max} = -75^\circ$ and $\theta_{max} = 75^\circ$. Each input entry to the SNN is a product of the measured ${\rm S}_2$ and the smoothed power signal for $m$ detectors. The smoothing is provided by a Savitzky–Golay filter (SGF).  The smoothing only occur over input angles since we assume that each output-angle detector is independent (that there is no signal processing coordinated between detectors except at the neural network). 

The SNN model is a two-layer SNN with nonlinear ``gelu'' activations and loss function that is the mean absolute error of $[\theta_i, {\rm S}_2, {\rm S}_3]$ predictions where each parameter has a range [-1, 1]. A validation split of 0.1 is used and the variance in the prediction is calculated for 15 test splits. 

\subsection{Details regarding Savitzky-Golay Filtering (SGF)}

The presence of the SGF mostly improves the pointing error prediction; the mean absolute error of ${\rm S}_2$ and ${\rm S}_3$ remain largely unchanged [Fig. \ref{fig:SSNIntuition}]. The disordered sample clearly achieves lower prediction errors. The pointing error drops when the window increases whereas the polarization information is constant. Random or disordered samples achieve lower MAE while the variance is larger.

\begin{figure}[htb!]
    \centering
    \includegraphics[width=.8\linewidth]{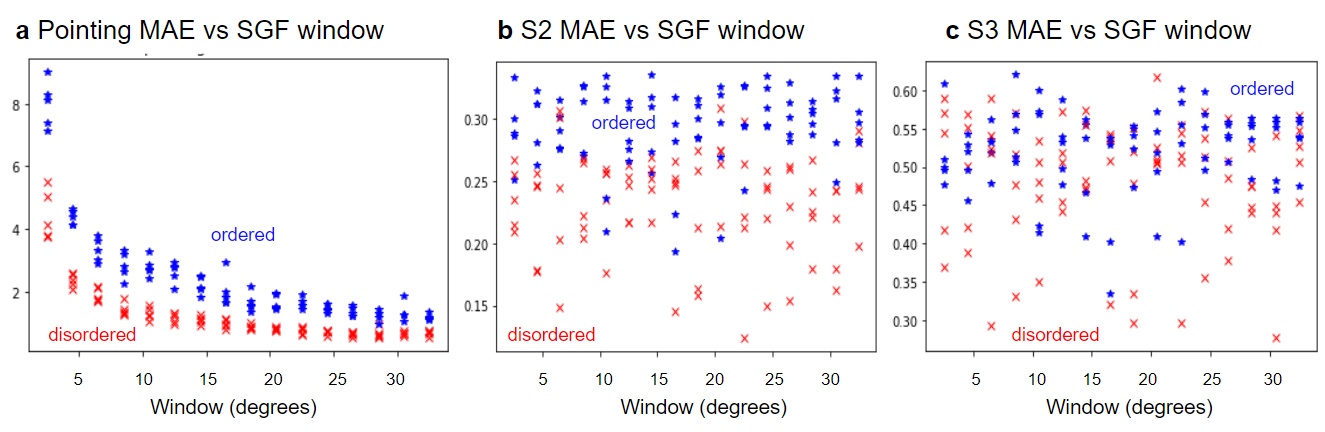}
    \caption{MAE for different Savitzky-Golay filter (SGF) windows. (a) Pointing MAE. (b) S$_2$ MAE. (c) S$_3$ MAE. The pointing error goes down with larger windowing whereas the polarization information is constant.  
}
    \label{fig:SSNIntuition}
\end{figure}

\subsection{Definition of compression ratio}
We define the compression ratio ${C_{\rm R}}$ as the improvement of the prediction error with encoding compared to the prediction error from conventional sampling \cite{Kilic2022}:
\begin{eqnarray}
{C_{\rm R}} &=& \frac{\varepsilon_{\rm Conventional}}{\varepsilon_{\rm Encoded}}\\
&\approx& \frac{\Delta\theta_o}{2\varepsilon_{\rm Pointing}}\frac{2(\pi- \varepsilon_{\rm S_2}\varepsilon_{\rm S_3})}{\pi} 
  \label{CR}
\end{eqnarray}
where $\Delta\theta_o$ is the average detector spacing, $\varepsilon_{\rm Pointing}$, $\varepsilon_{\rm S_2}$, and $\varepsilon_{\rm S_3}$ are the SNN prediction mean absolute error (MAE) associated with pointing, ${\rm S}_2$ and ${\rm S}_3$, respectively. The factors of $\pi$ in Eq. (\ref{CR}) refer to the area of the equatorial-plane unit circle of the Poincare sphere. Eq. (\ref{CR}) is an approximation, since we do not predict ${\rm S}_0$ and our inputs are ${\rm S}_2$, ${\rm S}_3=\pm 1$ while we consider a regression model where ${\rm S}_2$ and ${\rm S}_3$ are independent variables. For the purpose of this investigation, Eq. (\ref{CR}) is a reliable measure since ${C_{\rm R}}$ decreases when the polarization prediction error increases. Moreover, for a solid beam spot, the pointing MAE is approximately half of the detector spacing $\varepsilon_{\rm Pointing} \sim \Delta\theta_o/2$ and when $\varepsilon_{\rm S_2},\varepsilon_{\rm S_3}=0$, Eq. (\ref{CR}) yields ${C_{\rm R}} =2$. This factor of two is expected since the full-Stokes polarization characterization typically requires four separate measurements while our encoded approach employs only two.
\end{suppinfo}

\providecommand{\latin}[1]{#1}
\makeatletter
\providecommand{\doi}
  {\begingroup\let\do\@makeother\dospecials
  \catcode`\{=1 \catcode`\}=2 \doi@aux}
\providecommand{\doi@aux}[1]{\endgroup\texttt{#1}}
\makeatother
\providecommand*\mcitethebibliography{\thebibliography}
\csname @ifundefined\endcsname{endmcitethebibliography}  {\let\endmcitethebibliography\endthebibliography}{}

\end{document}